\documentclass[twocolumn,showpacs,preprintnumbers,pre,amsmath,amssymb]{revtex4}
\usepackage{amssymb}
\usepackage{amsbsy}
\usepackage{epsf}
\usepackage[cp866nav]{inputenc}
\usepackage{revsymb}
\usepackage[T2A,OT1]{fontenc}
\usepackage{graphicx}

\newcommand{\be}{\begin{equation}}
\newcommand{\ee}{\end{equation}}
\newcommand{\bea}{\begin{eqnarray}}
\newcommand{\eea}{\end{eqnarray}}

\renewcommand{\theequation}{\arabic{section}.\arabic{equation}}
\begin{document}

\title
{Coherence, decoherence, and memory effects in the problems of
quantum surface diffusion}
\author{V.V. Ignatyuk}
\affiliation{ Institute for Condensed Matter Physics, 1
Svientsitskii Street, 79011, Lviv, Ukraine}

\date{\today}

\begin{abstract}

{We consider surface diffusion of a single particle, which
performs site-to-site under-barrier hopping, fulfils intrasite
motion between the ground and the first excited states within a
quantum well, and interacts with surface phonons. On the basis of
quantum kinetic equations for one-particle distribution functions
obtained earlier we study the coherent and incoherent motion of
the adparticle. In the latter case we derive the generalized
diffusion coefficients and study various dynamic regimes of the
adparticle. The critical values of the coupling constant
$G_{cr}(T,\Omega)$, which separate domains with possible
recrossing from those with the monotonic motion of the adparticle,
are calculated as functions of a temperature $T$ and a vibrational
frequency $\Omega$. These domains are found to coincide with the
regions where the experimentally observed diffusion coefficients
change its behavior from weakly dependent on $T$ to quite a
sensitive function of the temperature. We also evaluate the
off-diagonal (relative to the site labels) distribution functions
both in the Markovian limit and when the memory effects become
important. The obtained results are discussed in the context of
the ``long tails'' problem of the generalized diffusion
coefficients, the recrossing/multiple crossing phenomena, and an
eventual interrelation between the adparticle  dynamics at short
times and the temperature dependence of the diffusion coefficients
measured experimentally.}

%\keywords  Quantum kinetic theory, models of surface chemical
%reactions, diffusion coefficient, rate constant

\pacs{05.60.-k, 63.20.K-, 66.10.cg, 68.35.Fx}
\end{abstract}

\maketitle

 \setcounter{equation}{0}

\section{Introduction}

A phenomenon of quantum diffusion of light particles
(mostly, hydrogen and its isotopes), adsorbed at solid
surfaces, in the last decades is a subject of joint efforts of
investigators in various fields of sciences. An attention of the
scientists is dictated by its relevance in the technological
processes like heterogeneous catalysis \cite{catalysis}, fuel cell
production \cite{q-jumps}, chemical reactions of hydrogen
transfer \cite{Htransfer}, and series of physical phenomena
occurring at the fluid-gas interfaces. On the other hand, the
diffusion of hydrogen is of fundamental interest from a basic
physical point of view, being a favorite system for theoretical
analysis \cite{PRB-41-1990,qkinet,Ferrando} and computer
simulations \cite{JJ-TCF,Japan}. Wide perspectives have been opened after creation of the scanning
tunnelling microscope (STM) \cite{single}, allowing to ``touch'' a
single molecule at the surface and to carry out experiments which
previously were only imagined. At the same time a rapid
development of powerful methods of computer experiments like
quantum molecular dynamics \cite{qmd} or Monte Carlo Wave Function
formalism \cite{36Ferrando} allows a direct analysis of the particle
trajectories in real space and time. Recent results showing that
even such the ``heavy'' atoms like Cr on the Au(111) surface
\cite{Ohresser} or Na on Cu(001) surfaces \cite{Casado} manifest a
great deal of the underbarrier tunnelling bring us to a conclusion
that a traditional viewpoint on the quantum diffusion as an
inherent feature of light particles only is far from reality,
and a fresh look at such processes is quite topical.

A task for the theorists - to create reliable models describing a
quantum surface diffusion and to develop effective methods of the
calculation of diffusion coefficients with taking into account all
the interactions (``adsorbate--adsorbate'',
``substrate--adsorbate'' etc.) has been accomplished to a great
extent. There should be mentioned the works
\cite{Kondo,PRL-62-1989,18PreprintMorozov} in which a bulk quantum
diffusion of light particles has been studied: its description
requires similar theoretical methods, and elaborated schemes can
be considered as starting points for investigation of the surface
quantum diffusion. They were followed by the papers where quantum
hopping was moved from the bulk to the surface
\cite{PRB-41-1990,qkinet,Ferrando,JCP1,JCP2,Zhu}. Generally, the
concept of a small polaron \cite{PRB-41-1990,JCP1,JCP2} or its
modification \cite{Ferrando} has been applied with going beyond
the linear ``adatom-phonon'' coupling to consider the anharmonic
terms in ``adsorbate--substrate'' interaction \cite{Zhu}. The
latter case together with consideration of electronic friction in
the system \cite{Kondo,Kagan} and direct ``adsorbate--adsorbate''
interaction \cite{Stasyuk} are very important, because they
provide additional channels of particle scattering and ensure
finite values of the diffusion coefficients \cite{MorozovBook}.

The diffusion coefficients are traditionally determined via the
Green-Kubo relations \cite{MorozovBook}, the low-frequency and
small wave-vector limit of the dynamic structure factor
\cite{Ferrando,PhysA195}, or zeroth moments of the ``flux-flux''
time correlation functions for classical
\cite{PhysA195,SurScience311} or quantum systems
\cite{JJ-TCF,Japan}. The latter approach is of particular interest
because it allows one not only to study in detail the well-known
``recrossing/miltiple crossing'' problem \cite{JJ-TCF,Japan}, but
also to derive a new quantum transition state theory \cite{TSWP}.

The quantum diffusion coefficient is known to consist of two terms
of a different physical origin. A coherent term \cite{JCP2,qkinet}
characterizes the way in which the dephasing limits the band
motion of the adatom by destruction of the coherence of the
hopping probabilities when the adatom--thermal bath coupling
induces random fluctuations of each phase. This term has a pure
quantum nature and is related to the competition between the
tunnel\-ling mechanism, which tends to preserve the coherence, and
the dephasing  which characterizes damping due to the scattering
process. The coherent part $D_{coh}(T)$ of the diffusion
coefficient weakly depends on temperature at low $T$ in contrast
to the incoherent one $D_{in}(T)$, which decreases to zero when
$T\to 0$. The incoherent contribution describes processes in which
the surface dynamics induces fluctuations of the tunnelling matrix
elements between two Wannier states allowing the adparticle to
perform a transition from one state to another by creating or
annihilating surface phonons. Though the temperature dependence of
the diffusion coefficients was studied profoundly for both small
polaron model \cite{JCP2} and its modification \cite{Ferrando},
some questions remain unclarified (especially, in the
weak-coupling limit, where the contributions of $D_{coh}(T)$ and
$D_{in}(T)$ are of the same order). The most challenging are the
following: i) a justification of the multiphonon expansion; ii) a
correct definition of the activation energy (e.g. when the main
contribution arises from the acoustic branch of phonon spectrum);
iii) an introduction of the additional channels of particles
scattering ensuring correct values of the diffusion coefficients;
iv) an influence of the memory effects on the temperature
dependence of the diffusion coefficients and on the short-time
dynamics of the adsorbate.

The latter problem should be considered from several standpoints.
First of all, a thorough study of the short-time dynamics of the
adparticle allows one to distinguish between various scenarios of the
adsorbate motion (a transition from the coherent to incoherent
regime, a presence of multiple or long hopping etc.) that
could be helpful for a deeper insight into the microscopic
picture of the process \cite{myPRE}. Secondly, we can answer the
question: what can enhance or suppress the particle motion. For
instance, an eventual recrossing reduces the values of the
diffusion coefficients, whereas a multiple crossing increases them.
At last, such a theoretical analysis can give some recommendations
for experimentalists how to perform evaluation of the
diffusion coefficients more effectively. It is known \cite{single}
that at temperatures above 80 K the diffusion rate of hydrogen
is too fast to be followed by the standard atom-tracking technique,
while below 50 K the opposite problem occurs, and it is necessary
to minimize the influence of the STM tip on the adsorbate due to a
prolonged interaction. Thus, it would be tempting to relate a
change in the character of the short-time dynamics of the
adsorbate to possible crossover from one typical temperature
behavior of the diffusion coefficients to another and to give a
prognosis about the $T$-dependence of the diffusion coefficient having
only an information about the adatom dynamics at the initial stage
of its motion.

In the present paper, which is a logical continuation of Ref.~\cite{myPRE}, we try to give an answer to the
question about the interrelation between the process of the
decoherence in the ``adsorbate--substrate'' system (leading to the
dissipative dynamics of the adsorbate) and the memory effects.
Here we also use the method of the quantum kinetic
equations \cite{MorozovBook,cmp2004,JCP2}.

The subject of our study is a single adsorbate which performs an
underbarrier hopping to the nearest adsorption sites, fulfils
intrasite motion between two different quantum states within a
quantum well, and interacts with acoustic phonons. We define the
conditions under which the adatom dynamics is definitely nonlocal
in time, derive expressions for the generalized
(time-dependent) diffusion coefficients, and study the influence of their
long-time asymptotics on the temperature behavior of the
transport coefficients measured experimentally.
A particular emphasis is put on the critical regimes separating
dynamics with mainly coherent contribution to the diffusion
coefficient from the dominance of that of the incoherent origin.
We show that a transition domain from the oscillating dynamics of
the adparticle to the monotonic motion subject to the critical
coupling constant $G_{cr}(T,\Omega)$ falls perfectly into the same
region of $G_{cr}(T,\Omega)$ values as a crossover of the
temperature dependence of the diffusion coefficients from weakly
dependent on $T$ to quite a sensitive function of the temperature.

A special attention is paid to evaluation of the off-diagonal
(relative to the site labels) distribution functions, which
describe how fast the loss of the coherence occurs. While the
study of the generalized diffusion coefficients (being related to
the ``velocity--velocity'' autocorrelation function) allows one to
make a conclusion about the recrossing phenomenon, behavior of
the off-diagonal distribution functions shows us how a multiple
crossing of the dividing surfaces (placed at the neighboring
adsorption sites) by a moving adparticle can proceed.

Our paper is organized in the following way. In Section II we
start from the unitary transformed Hamiltonian for a dissipative
two-level system on a new correlated basis
\cite{JCP1,JCP2,inclusion}. In Section III using
the obtained earlier \cite{myPRE} non-Markovian equations for non-equilibrium distribution
functions, we investigate a long-time
asymptotics of the kinetic kernels determining a dissipative
motion of the adsorbate. A particular case of a completely
coherent dynamics, which corresponds either to $T\to 0$ or zero
coupling limits, is considered. The expression for the generalized
diffusion coefficients is obtained in the next Section. In Section
V the Markovian approximation for these functions is considered, and
a temperature behavior of the experimentally observed diffusion
coefficient is studied in a weak-coupling limit. In Section VI a
thorough analysis of the critical diagrams separating different
kinds of the adparticle motion is performed, and the
interconnection of the obtained results with those of the previous
Section is established. In Section VII the off-diagonal
non-equilibrium distribution function is evaluated at different
values of the tunnelling constant at assumption of a continuous
media; the obtained results are considered in the context of the
multiple crossing, and a conclusion about validity of the
Markovian approximation is inferred. In the last Section we
discuss briefly the obtained results and draw final conclusions.

\setcounter{equation}{0}
\section{Unitary transformed Hamiltonian of the ``substrate--adsorbate'' system}

To specify all interactions in the ``substrate--adsorbate'' system
we choose a Hamiltonian, which allows site-to-site tunnelling of
the adsorbate, intrasite oscillations of the adparticle between the
ground and the 1-st excited states within the potential well, and
interaction of the adparticles with the lattice (both by density and oscillation modes). We use the same
basic Hamiltonian as in Ref.~\cite{JCP1}.

Usually, in quantum diffusion problems one can consider the
``substrate--adsorbate'' coupling to be arbitrary. On the other
hand, one-particle characteristics of the system dealt with the
intersite hopping and the intrasite motion are treated as small
parameters. In such a case it is useful \cite{inclusion,JCP1,JCP2}
to start from the unitary transformed Hamiltonian on a new
correlated basis, which provides a better zeroth-order
representation: the sequence of unitary transformations has the
effect of changing to a representation in which the adsorbate is
localized at the left ($L$-state) or at the right ($R$-state) end
of an adsorption site, and in which there is a correlated
displacement of the lattice. Thus, a starting point in our study
is the transformed Hamiltonian of the system of adparticles
\cite{JCP1}: \be\label{Ht} \tilde
H=H_{intra}+H_{inter}+H_{pp}+H_B\equiv H'+H_{pp}+H_B. \ee The term
$H_{intra}$ describes the lattice-modified \underline{intrasite}
dynamics of the adparticle: \be\label{Hs} H_{intra}\!=\!\sum_s
\frac{U}{2}n_s
(n_s-1)-\!\left(\frac{\hbar\Omega}{2}B_{s}a^{\dagger}_{s L}a_{s
R}+\mbox{h.c.}\right),\ee \vspace*{-2mm} \be\label{Bs}
B_{s}=\exp\left[-2 \sum_{q}\frac{\chi_{s
q}}{\hbar\omega_q}(b_q-b^{\dagger}_q)\right]\ee
with a lattice induced operator exponent $B_{s}$, where denotation
h.c. means Hermitian conjugation. The other denotations in
Eqs.~(\ref{Hs})-(\ref{Bs}) are the following: $U$ means the on-site
Hubbard repulsion; $\Omega$ stands for the vibrational frequency
between $L$- and $R$-states with the corresponding creation
$a^{\dagger}_{sL}$, $a^{\dagger}_{sR}$ or annihilation $a_{sL}$,
$a_{sR}$ operators of the adparticle at a given site $s$ (in these
notations $n_s=a^{\dagger}_{sL}a_{sL}+a^{\dagger}_{sR}a_{sR}$). It
has to be stressed that $\hbar\Omega$ coincides with the
oscillation energy between the ground and the 1-st excited
vibrational states at the assumption that zero of energy lies
midway between these two levels \cite{JCP1,JCP2}. The strength
$\chi_{s q}$ describes coupling of phonons with the energy
$H_B=\sum_q\hbar \omega_q b^{\dagger}_q b_q$ to the oscillation
modes of the adsorbate, and only an acoustic branch $\omega_q$ of
the substrate motion is taken into account.

The second summand in (\ref{Ht}) \bea\nonumber \label{HT}
&&H_{inter}\!=\sum_{\langle ss'\rangle}\!\! t_{ch}(B^{LR}_{s
s'}a^{\dagger}_{s L}a_{s' R}+B^{RL}_{s
s'}a^{\dagger}_{s R}a_{s' L})\\
&&  +t_{pr}(B^{LL}_{s s'}a^{\dagger}_{s L}a_{s' L}+B^{RR}_{s
s'}a^{\dagger}_{s R}a_{s' R})
 \eea
is the \underline{intersite} tunnelling term  with end-changing
(end-preserving) amplitudes
 $t_{ch}$ ($t_{pr}$), which are the linear
 combinations
$t_{ch\atop pr}=\frac{1}{2} (t_{1}\pm t_{0})$
 of the initial tunnelling amplitudes (usually, a condition $t_1\gg
 t_0$ is valid, thus an underbarrier hopping of the particle is
 determined by the tunnelling constant $t_1$ between the excited
 states rather than  between the ground states). An abbreviation $\langle ss'\rangle$ in (\ref{HT}) denotes a sum
over the nearest-neighbor sites.

The structure of lattice induced operators $B$ in Eq.~(\ref{HT}) is similar to that of
(\ref{Bs}): \bea\nonumber \label{Btun}B^{LR, RL}_{s
s'}=\exp[-\!\sum_q(\Delta_q^{s
s'}\!\pm\!^{(+)}\!\delta_{q}^{s s'})(b_q-b^{\dagger}_q)],\\
B^{LL,  RR}_{s s'}=\exp[-\!\sum_q(\Delta_q^{s
s'}\!\pm\!^{(-)}\!\delta_q^{s s'})(b_q-b^{\dagger}_q)],\\
%\be\label{deltas}
\nonumber \Delta_q^{s s'}=\frac{\gamma_{s q}-\gamma_{s'
q}}{\hbar\omega_q},\qquad ^{(\pm)}\delta_q^{s s'}=\frac{\chi_{s
q}\pm\chi_{s' q}}{\hbar\omega_q},
 %\ee
 \eea
where the upper sign corresponds to the first superscript, and the
strength $\gamma_{s q}$ describes coupling of phonons with the density
mode of the adsorbate. In a 1$D$ case both $\gamma_{s q}$ and
$\chi_{s q}$ can be expressed explicitly via the lattice
parameters \cite{JCP1}; for a 2$D$ infinite lattice we shall introduce
the lattice spectral weight functions \cite{JCP1,JCP2} to
describe ``substrate-adsorbate'' interaction.

The last but one term of the Hamiltonian (\ref{Ht})
 \bea\nonumber\label{Hpp}
&&H_{pp}=-\sum_{\langle
ss'\rangle}\left\{\!C^{DD}_{s,s'}n_{s}n_{s'}+2
C^{DO}_{s,s'}n_{s}(n_{s'
L}-n_{s' R})\right.\\
&&\left.+C^{OO}_{s,s'}(n_{s L}-n_{s R})(n_{s' L}-n_{s' R})\right\}
 \eea
 describes the \underline{particle--particle lattice induced
 interaction}. For the explicit expressions of the strengths
 $C^{ij}_{s,s'}$, $i,j=\{O,D\}$, and explanation of their physical meaning the reader is referred to
 Refs.~\cite{JCP1,JCP2,myPRE}. It has also to be stressed that a direct ``adsorbate--adsorbate''
interaction can be introduced at this stage, and it will modify the
expressions for $C^{ij}_{s,s'}$ to a certain extent.

To describe the ``substrate-adsorbate'' interaction we consider
site-independent end-changing spectral weight functions:
\be\label{Jintra} J(\omega)=\sum\limits_q\chi^2_{s
q}\delta(\omega-\omega_q), \ee \vspace*{-5mm}
%\label{Jintra}\label{Jc}\label{Jp}
 \bea\label{Jc}\nonumber
J_{LR}(\omega)\!=\!\!\sum\limits_q\left[(\gamma_{s q}-\gamma_{s'
q})\!+\!(\chi_{s q}+\chi_{s' q})
\right]^2\!\delta(\omega-\omega_q),
\\
\\
\nonumber J_{RL}(\omega)\!=\!\!\sum\limits_q\left[(\gamma_{s
q}-\gamma_{s' q})\!-\!(\chi_{s q}+\chi_{s' q})
\right]^2\!\delta(\omega-\omega_q),
 \eea
and end-preserving ones:
\bea\label{Jp}\nonumber
J_{LL}(\omega)\!=\!\!\sum\limits_q\left[(\gamma_{s q}-\gamma_{s'
q})\!+\!(\chi_{s q}-\chi_{s' q})
\right]^2\!\delta(\omega-\omega_q),
\\
\\
\nonumber
J_{RR}(\omega)\nonumber\!=\!\!\sum\limits_q\left[(\gamma_{s
q}-\gamma_{s' q})\!-\!(\chi_{s q}-\chi_{s' q})
\right]^2\!\delta(\omega-\omega_q).
 \eea
The function (\ref{Jintra}) describes the intrasite dynamics; the
functions (\ref{Jc}) are related to the intersite end-changing
processes, while (\ref{Jp}) are dealt with the intersite
end-preserving processes. At low frequencies the end-changing
spectral weight functions (labelled by the subscript $c$) are
approximately given by \be\label{Jc1} J_c(\omega)\approx\Biggl\{
\begin{array}{c}
\! 0,\qquad\qquad\omega <\omega_0,\\
\eta_c\omega^{D-2},\quad\omega>\omega_0,
\end{array}\ee
and the end-preserving (with the subscript $p$) ones by
\be\label{Jp1} J_p(\omega)\approx\Biggl\{
\begin{array}{c}
\! 0,\qquad\quad\omega <\omega_0,\\
\eta_p\omega^{D},\quad\omega>\omega_0
\end{array}\ee
with \be\label{GG}\eta_c=10 G,\qquad\qquad \eta_p=12.5 G,\ee given
in units of the dimensionless coupling constant \cite{JCP1}
\be\label{Gamma} G=\frac{\Gamma^2}{M\omega^3_{max}}.\ee In
Eqs.~(\ref{Jc1})-(\ref{Gamma}) $D$ labels dimensionality of the
lattice; $M$ denotes the mass of a substrate atom; $\omega_{max}$
stands for the Debye frequency, and the coupling strength $\Gamma$
is expressed via the mean value of the distortion potential over the
localized Wannier states.

It is seen from  (\ref{Jc1})-(\ref{Jp1}) that the lattice is
allowed to possess a nonzero lowest frequency $\omega_0$. At first
glance, the presence of a gap in the spectrum of acoustic phonons
on the infinite lattice seems to be rather controversial. There
were several attempts \cite{Leggett,18PreprintMorozov,JCP1} to
associate the value $\omega^{-1}_0$ with a finiteness of the
particle life-time $\tau_{0}\gg\omega_{max}^{-1}$ in a quantum
well. We are not going to repeat in our article the main
reasonings of cited papers, noting that the obtained
results for the surface quantum diffusion coefficients are quite
insensitive to the value of $\omega_0$, provided the temperature
is much higher than $\hbar\omega_0/k_B$ \cite{JCP2,myPRE}.

Another interesting feature of the low-frequency dependence of the
spectral weight functions (\ref{Jc1})-(\ref{Jp1}) is that they are
scaled like in the case of a bulk diffusion of a light inclusion,
assisted by optical phonons \cite{MorozovBook,inclusion}. The only
difference is the value of the parameter $D$, which is found to be
equal to 5.

Now let us proceed to the investigation of the adsorbate dynamics
using the system of quantum kinetic equations.

\setcounter{equation}{0}
\section{Kinetic equations for one-particle non-equilibrium functions of adsorbate}

The system of quantum kinetic equations for one-particle
non-equilibrium functions of the adsorbate can be obtained using the
equation for the reduced density matrix $\rho_S(t)$
\cite{MorozovBook}. Considering the first two terms of the
 Hamiltonian (\ref{Ht}) as a small per\-tur\-ba\-tion, we can
 construct a closed systems of kinetic equations up to the second order in
 $H'$. These equations turn out to be nonlocal in time, so it is
 convenient to perform a Laplace transformation for the diagonal
$f_{s,s}(t)=\sum\limits_{i=L,R}\langle a^{\dagger}_{s i} a_{s
i}\rangle^t_S$ and off-diagonal
$f_{s,s'}(t)=\sum\limits_{i=L,R}\langle a^{\dagger}_{s' i} a_{s
i}\rangle^t_S$ one-particle non-equilibrium distribution
functions. In the single particle limit we obtain \cite{myPRE} the
following chain of equations for the Laplace transforms
$\tilde{f}_{s,s}(z)$, $\tilde{f}_{s,s+n}(z)$ (the index $s+n$ means
the nearest neighboring site with respect to $s$):

\bea\label{fss}\nonumber
&&z\tilde{f}_{s,s}(z)\!-\!f_{s,s}(t\!=\!0)\!=\!-\frac{i}{\hbar}
t_{inter}\!\!\sum\limits_n(\tilde{f}_{s,s+n}(z)-\tilde{f}_{s+n,
s}(z))\\
&&
-\tilde{\gamma}_{inter}(z)\left(2\tilde{f}_{s,s}(z)-\sum\limits_n\tilde{f}_{s+n,
s+n}(z)\right),\\\nonumber
&&z\tilde{f}_{s,s+n}(z)-\!f_{s,s+n}(t\!=\!0)\!=\!-\frac{i}{\hbar}
t_{inter}\left(\tilde{f}_{s+n,s+n}(z)-\tilde{f}_{s,
s}(z)\right. \\
\label{fss1} &&\left.+\tilde{f}_{s-n,s+n}(z)-\tilde{f}_{s,s+2n}(z)\right)\\
\nonumber &&
-\left(\tilde{\gamma}_{inter}(z)\!+\!\tilde{\gamma}_{intra}(z)\right)\tilde{f}_{s,s+n}(z)+\tilde{\gamma}^{+}_{LL}(z)\tilde{f}_{s+n,
s}(z).\!\! \eea A similar chain of equations has been obtained in
Ref.~\cite{JCP2}, but the authors applied the Markovian
approximation for the kinetic kernels and did not study the
short-\-time dynamics of the adsorbate.

Linearity of Eqs. (\ref{fss})-(\ref{fss1}) is the result of the
single-particle approximation, and the question about statistics
of the adparticle loses its significance. Note, that a similar
linear approximation could be performed also at low coverage of
the adsorbate. However, at low-to-moderate coverage one has to
retain all nonlinear terms and at high coverage to include
non-equilibrium correlation functions into the set of dynamical
variables of the abbreviated description \cite{MorozovBook}.

Some words have to be said about all constituents of the kinetic
equations (\ref{fss})-(\ref{fss1}). The first terms of r.h.s.
describe a nondissipative coherent motion of the adsorbate with
the renormalized tunnelling amplitude
$$t_{inter}=t_{pr}\exp\left[-\frac{1}{2}\int\limits^1_{\omega_0}
d\omega\frac{J_p(\omega)}{\omega^2}\coth
\left(\frac{\hbar\omega}{2k_B T} \right) \right]
$$\vspace{-0.5cm}
\be\label{Tint}= t_{pr}\omega_0^{\eta_p k_B T}\left[\eta_p k_B
T\sinh\left(\frac{1}{2\eta_p k_B T} \right)\right]^{-\eta_p k_B
T}\!.\ee In fact, $t_{inter}$ corresponds to the polaron band
narrowing due to the ``substrate--adsorbate'' interaction. Hereafter
we use dimensionless frequencies in the units of $\omega_{max}$
and temperatures in the units of $\hbar\omega_{max}/k_B$.

The kinetic kernel \be\label{GamInter}
\tilde{\gamma}_{inter}(z)=4\tilde{\gamma}_{LL}(z)+2\tilde{\gamma}_{LR}(z)+2\tilde{\gamma}_{RL}(z)\ee
corresponds to the dissipative intersite motion of the adsorbate
and describes processes, when the adparticle performs series of
random site-to-site hoppings (with or without the change of its
quantum states) owing to the interaction with a bath. The kinetic
kernel $\tilde{\gamma}_{intra}(z)$ in Eq.~(\ref{fss1}) describes a
dissipative intrasite dynamics, when the adsorbate during its
scattering from the lattice gets enough energy from the bath to be
excited from the ground state to the upper level within one
adsorption site (the opposite process of particle de-excitation
with a phonon emission is also taken into consideration). The
rates $\tilde{\gamma}_{intra}(z)$, $\tilde{\gamma}_{inter}(z)$ can
be obtained from the Laplace transformation of the kinetic kernels
\[ \gamma_x(\tau)=\omega_{max}\lambda_x^2 J_0^4\left(
\frac{2t_{inter}\tau}{\hbar}\right)\vspace{-3mm}
\]
\be\label{Gamx} \times\mbox{Re}\{\exp[
-(\varphi_x(0)-\varphi_x(\tau))]-\exp[-\varphi_x(0)]\}, \ee
%\ee
%%%
\[\gamma_{LL}^+(\tau)\!=\omega_{max}t_{pr}^2
J_0^4\left( \frac{2t_{inter}\tau}{\hbar}\right)
\]\vspace{-5mm}
\be\label{GamLLPlus}\times\mbox{Re}\{\exp[
-(\varphi_{LL}(0)+\varphi_{LL}(\tau))]-\exp[-\varphi_{LL}(0)]\},\ee
where \be\label{phi}\hspace*{-5mm}
\varphi_x(\tau)\!=\!\!\!\int\limits_{\omega_0}^1\!\!\frac{J_x(\omega)}{\omega^2}\!\left[\coth\!\left(
\frac{\hbar\omega}{2 k_B T}\!\right)\!\cos(\omega
\tau)-i\sin(\omega \tau) \!\right].\ee In Table I we present the
amplitudes $\lambda_x$ and the spectral weight functions
$J_x(\omega)$, relevant to the rates
(\ref{Gamx})-(\ref{GamLLPlus}) appearing in the kinetic equations
(\ref{fss})-(\ref{fss1}).
\begin{table}[htb]
\caption{Rates $\gamma_x$, dimensionless amplitudes $\lambda_x$,
and spectral weight functions $J_x(\omega)$ along with their
low-frequency limits (\ref{Jc1})-(\ref{Jp1}).}
\begin{center}
\begin{ruledtabular}\begin{tabular}{ccc}
 $\gamma_x$ & $\lambda_x$ & $
     J_x(\omega)$ \\
     \hline  $\gamma_{intra}$ & $\Omega/2\omega_{max}$ &$J_{intra}(\omega)=J_{c}(\omega)$\\
$\gamma_{LR}$ & $ -(t_1+t_0)/2\hbar\omega_{max}$ & $J_{LR}(\omega)=J_{c}(\omega)$  \\
$\gamma_{RL}$ & $ -(t_1+t_0)/2\hbar\omega_{max}$ & $J_{RL}(\omega)=J_{c}(\omega)$  \\
$\gamma_{LL}$ & $ (t_1-t_0)/2\hbar\omega_{max}$ & $J_{LL}(\omega)=J_{p}(\omega)$  \\
$\gamma_{RR}$ & $ (t_1-t_0)/2\hbar\omega_{max}$ & $J_{RR}(\omega)=J_{p}(\omega)$  \\

 \end{tabular}\end{ruledtabular}
     \label{Jx}
     \end{center}
     \vspace{-6mm}\end{table}
\noindent The functions (\ref{phi}) yield lattice contributions to
the kinetic kernels, and the zeroth order Bessel function $J_0$ in
(\ref{Gamx})-(\ref{GamLLPlus}) accounts for the particle
contribution. The latter function ensures a convergence of the
time integrals of (\ref{Gamx})-(\ref{GamLLPlus}) at an arbitrary
value of the coupling constant, though from the mathematical point
of view it exceeds the required accuracy, being higher than of the
second order in tunnelling amplitudes. This result is known to
follow from going beyond the limits of the 2-nd Born approximation
for the kinetic kernels \cite{MorozovBook}, and provides a
relaxation of the kernels to zero when $t\to\infty$. However, one
has to be careful when dealing with short-time dynamics of the
system, because in spite of the kinetic kernels decay the problem
of energy conservation in the system appears \cite{cmp2004}. Due
to this reason we will omit the Bessel functions in the
expressions (\ref{Gamx})-(\ref{GamLLPlus}), using instead a
concept of the finite life-time $\tau_{0}=\omega^{-1}_0$ of the
adparticle at a given adsorption cite.

A study of the long-time asymptotics of the kinetic kernels allows
one to establish a one-to-one correspondence between the
low-frequency behavior of the spectral weight functions
$J(\omega)$ and a damping of the kernels
(\ref{Gamx})-(\ref{GamLLPlus}) at long times. These results are
summarized in Table II, where the constants $a_i$,
$i=\{0,\ldots,3\}$, are introduced just to describe a particular time
behavior of $\gamma(\tau)$ (in general case, these values are
defined by the system parameters).
\begin{table}[htb]
\caption{Relation between a low-frequency asymptotics of the
spectral weight functions and a long-time relaxation of the
kinetic kernels.}
\begin{center}
\begin{ruledtabular}\begin{tabular}{cc}
 $J(\omega)\sim$ & $\gamma(\tau)\sim$\\
     \hline  $\omega^0$ & $\exp(-a_0\tau^2)$\\
$\omega^1$ & $ \exp(-a_1\tau)$\\
$\omega^2$ & $1/\tau^{a_2}$\\
\quad $\omega^n$, $n\ge 3$ & $ \exp(-a_3)$ \\

 \end{tabular}\end{ruledtabular}
     \label{JwGam-tau}
     \end{center}
     \vspace{-6mm}\end{table}
\noindent It is seen from Table II that we pass from a fast
relaxation of the kinetic kernels at $J(\omega)\sim\omega^n$,
$\{n=0,1\}$, through long tails at $n=2$ to the divergent
transport coefficients  (which are defined via time integrals of
(\ref{Gamx})-(\ref{GamLLPlus})) at $n=3$. Note, that
the system dimensionality, as it is seen from
Eqs.~(\ref{Jc1})-(\ref{Jp1}), influences the long-time asymptotics
of the kinetic kernels above all: in the bulk the only linear
``substrate--adsorbate'' coupling does not ensure the finite values
of the diffusion coefficients. As it has been already mentioned,
this divergency can be eliminated by introduction of the
additional channels of the particle scattering (electronic
friction \cite{Kondo,Kagan}, anharmonic terms in
``adsorbate--substrate'' interaction \cite{18PreprintMorozov,Zhu}
or direct ``adsorbate--adsorbate'' interaction), which changes the
low-frequency asymptotics of the spectral weight functions. For
instance, taking into account the first two factors yields damping
of the kinetic kernels  as $\exp(-a_1 t)$
\cite{Kondo,Kagan,18PreprintMorozov}.

Keeping in mind the data of Table II, it is possible to evaluate the
long-time asymptotics of the end-changing (see first three rows of
Table I) and end-preserving (last two rows) kinetic rates
$\gamma_{ch\atop pr}(\tau)$. Both rates in the strong-coupling
limit ($G\ge 0.1$) decay as Gaussian functions
\be\label{PhicApprox}
\gamma_c(\tau)\sim\exp\left[-\eta_c|\ln\omega_0|\left(k_B
T\tau^2+i\tau \right) \right],\ee
\be\label{PhipApprox}\gamma_p(\tau)\sim\exp\left[-\frac{\eta_p}{2}
\left(k_B T\tau^2+i\tau \right)\right],\ee while at the
weak-coupling limit ($G\le 0.01$) the first of them still decays
as a Gaussian function, but the second of them behaves as

\be\label{PhipApprox1} \gamma_p(\tau)\sim 1/\tau^{A\eta_p k_B
T}-\omega_0^{2\eta_p k_B T},\qquad A=2.15.\ee

We will return to the expressions
(\ref{PhicApprox})-(\ref{PhipApprox1}) in Section V when
evaluating a temperature behavior of the diffusion coefficients.
We also leave aside the general cases of arbitrary $n$ (both the
sub-ohmic, $ 0\le n < 1$, and the super-ohmic, $n>1$, regimes) in
the low-frequency asymptotics of the  spectral weight functions,
referring readers to the review \cite{Leggett}. Instead, we
consider a limiting case of the vanishing coupling constant $G\to
0$, or, equivalently, the low-temperature limit $T\to 0$, when the
adparticle motion becomes completely coherent. For simplicity, we
limit ourselves to the 1$D$, single-band ($\Omega=0$)
approximation. In such a case the kinetic equations
(\ref{fss})-(\ref{fss1}) can be rewritten as \be\label{fssCoh}
\dot f_{s,s}(t)=-\frac{i}{\hbar} t_0\sum_n\left(
f_{s,s+n}(t)-f_{s+n,s}(t)\right), \ee \be\label{fssCoh1} \dot
f_{s,s+n}(t)\!\!=\!\!-\frac{i}{\hbar}t_0\left(
f_{s+n,s+n}(t)\!\!-\!\!f_{s,s}(t)\!\!+\!\!f_{s-n,s+n}(t)\!\!-\!\!f_{s,s+2n}(t)\right).
\ee  It is seen that in the site representation we face with a
coupled chain of equations involving all the lattice labels $s$.
However, this system of equations can be solved in the wave-vector
representation (the details of the solution can be found in
Appendix), giving the final result via the $s$-th order Bessel
functions $J_s$ as \bea\label{diagonal}
&&f_{s,s}(\tau)=J_s^2(2t_0\hbar^{-1}\tau), \eea
\bea\label{nondiagonal}
&&f_{s,s+1}(\tau)=\mbox{Re}\left[f_{s,s+1}(0)\right]+i
J_s(2t_0\hbar^{-1}\tau)J_{s+1}(2t_0\hbar^{-1}\tau). \eea The real
part of the off-diagonal distribution function can be evaluated by
the methods of equilibrium statistical mechanics. It defines a
strength of the transition, and does not evolve in time. Instead,
the imaginary part of the $f_{s,s+n}(\tau)$ is shifted by the
quarter-period with respect to the diagonal distribution functions
(see Fig.~1).
\begin{figure}[htb]
\centerline{\includegraphics[height=0.24\textheight]{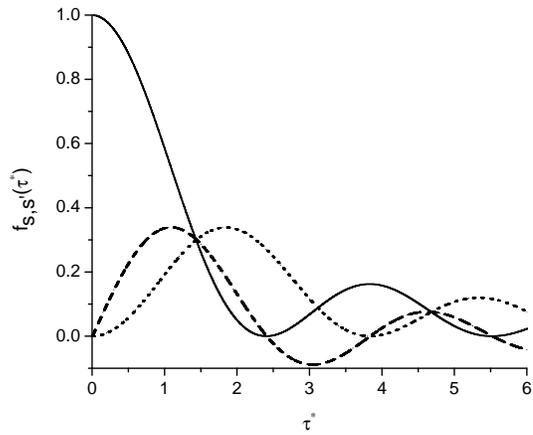}}
\vspace{0.2cm} \caption{The time dependence of the one-particle
nonequilibrium distribution functions $f_{0,0}(\tau^*)$ (solid
line), $f_{1,1}(\tau^*)$ (dotted line), and imaginary part of
$f_{0,1}(\tau^*)$ (dashed line), given by
Eqs.~(\ref{diagonal})-(\ref{nondiagonal}). The symbol $\tau^*$
denotes a time in the units $\hbar/(2 t_0)$.} \label{coherent}
\end{figure}
A particle, initially located, say, at $s=0$, starts its motion
toward the nearest adsorption site $s=1$. At that time the
probability to find the adparticle at the site $s=0$ reduces, the
probability to meet the adparticle at the site $s=1$ increases.
The imaginary part of the transition probability $f_{s,s+1}(\tau)$
reaches its maximum value when the increase/decrease of the diagonal
distribution functions becomes maximal ($\tau^*\approx 1$). When
the particle reaches the site $s=1$ at $\tau^*=2$, and the value
of $f_{1,1}(\tau)$ is maximal, the inverse motion of the part of the
wave packet towards the site $s=0$ begins, giving the negative
branch of $\mbox{Im}[f_{0,1}(\tau)]$ at the subsequent period of time.
Another part of the wave packet keeps on a motion towards the site
$s=2$ yeilding a positive branch of $\mbox{Re}[f_{0,2}(\tau)]$ and
$\mbox{Re}[f_{1,2}(\tau)]$ (not presented in Fig.~1).

In contrast to the classical picture, when a free particle
performs quasi-continuous  motion over the barriers, the reason
for oscillations of the distributions functions is purely quantum
mechanical: a superposition of the wave packets reflection from
the potential barriers and tunnelling through them.

To study the incoherent motion of adsorbate we have to obtain
expressions for the generalized diffusion coefficients. This task
is accomplished in the next Section.

\section{Generalized diffusion equation for the adparticle at the
surface}
\setcounter{equation}{0}

Our further advance is in the manner of the generalized collective
modes approach \cite{GCM1,GCM4}. To find the generalized diffusion
coefficient let us solve Eq.~(\ref{fss1}) with respect to the
hopping probabilities $\tilde{f}_{s,s+n}(z)$ and insert the
obtained result into Eq.~(\ref{fss}). After grouping the terms,
one obtains the following equation: \bea\nonumber\label{difEqn}
&&z\tilde{f}_{s,s}(z)-f_{s,s}(0)=\Biggl\{ \frac{2 t_{inter}^2
\hbar^{-2}}{z+\tilde{\gamma}_{inter}(z)+\tilde{\gamma}_{intra}(z)+\tilde{\gamma}_{LL}^+(z)}
\\
%\vspace{-0.5cm}\be\label{difEqn}\hspace*{7mm}
&&+\tilde{\gamma}_{inter}(z)\Biggr\}\left(\sum\limits_n
\tilde{f}_{s+n,s+n}(z)-2\tilde{f}_{s,s}(z)\right)\\
\nonumber
&&+\frac{t_{inter}^2 \hbar^{-2}}{z+\tilde{\gamma}_{inter}(z)+\tilde{\gamma}_{intra}(z)+\tilde{\gamma}_{LL}^+(z)}\\
\nonumber &&\times \mbox{Re}\Biggl( 2\tilde f_{s-n,s+n}(z)-\tilde
f_{s,s+2n}(z)+\tilde f_{s,s-2n}(z) \Biggr).\eea The ratio in the
braces describes a coherent contribution $\tilde{D}_{coh}(z)=2
t^2_{inter}\hbar^{-2}/\tilde{\gamma}_{total}(z)$ to the
generalized diffusion coefficient and can be interpreted
\cite{JCP2,myPRE} in terms of a simple model of band-type motion
limited by scattering from the lattice at temperatures large
relative to the bandwidth: \be\label{DDcoh1} \tilde D_{coh}(0)\sim
v^2/\tilde{\gamma}_{total}(0), \ee where $v=a t_{inter}/\hbar$
denotes the average velocity of the adsorbate, $a$ stands for a
substrate interatomic spacing, and
\be\label{Gamtotal}\tilde{\gamma}_{total}(z)=\tilde{\gamma}_{inter}(z)+\tilde{\gamma}_{intra}(z)+\tilde{\gamma}_{LL}^+(z)\ee
means the total rate of scattering from the lattice. A coherent
contribution $\tilde D_{coh}(z)$ characterizes the way in which
the dephasing limits the band motion of the adatom by destruction
of the coherence of the hopping probabilities $\tilde
f_{s,s+n}(z)$. Whereas the eigenstate of a free particle on the
surface is described by a superimposition of localized Wannier
states (this limiting case corresponds to the ballistic regime of
motion), the coupling with the thermal bath induces random
fluctuations of each phase which destroys the coherence of the
state.

The second term $\tilde{D}_{in}(z)\equiv\tilde{\gamma}_{inter}(z)$
in the braces of Eq.~(\ref{difEqn}) is an incoherent contribution
to the generalized diffusion coefficient. This is the result
expected from the random walk model for diffusion with
site-to-site hopping rate $\tilde{\gamma}_{inter}(z)$, describing
processes of the surface phonon creation/annihilation when the
particle performs a transition from one Wannier state to another.

The last term of the r.h.s of Eq.~(\ref{difEqn}) involves the
transition probabilities of the particle to perform a long hopping
with $|s-s'|>a$. In the case of consideration, when
$t_1\ll\hbar\Omega$, this term is very small, being of the 4-th
order in tunnelling amplitude $t_1$. If $t_1\sim\hbar\Omega$, it
is comparable with the second term of Eq.~(\ref{difEqn}), being
$\sim t_1^2$. Whatever the case, as it is shown in Appendix, the
last term does not contribute to the overall diffusion
coefficient.

Let us remind that in Section III we assumed the kinetic kernels
to be independent of the site label $s$. This assumption leads to
the absence of spatial non-locality in the expressions for
generalized diffusion coefficients, so the memory effects only are
taken into consideration.  A spatial inhomogeneity is the subject
of separate studies (see, e.g. Ref.~\cite{Chvoj}). The most
general case of the wave-vector dependent diffusion coefficients
also with the time non-locality, that allows one to get a deeper
insight into dynamics of the system at various time-spatial
scales, is a challenging topic of non-equilibrium surface
diffusion theory but lies beyond the scope of the present paper.

A multiplier $\sum\limits_n
\tilde{f}_{s+n,s+n}(z)-2\tilde{f}_{s,s}(z)$ at the braces in
Eq.~(\ref{difEqn}) in the continuous media limit, when the
interatomic spacing tends to zero, converts to the second
derivative with respect to the space variable (for 1$D$ lattice)
times $a^2$ or to the Laplace operator (for 2$D$ lattice in
absence of the next-to-nearest-neighbor hopping) times $4a^2$. It
was shown in Ref.~\cite{myPRE} that in the continuous media limit,
and when coupling between the adparticle and the surface is strong
enough, it is possible to obtain the Telegrapher's equation for
the nonequilibrium distribution function $n(r,t)$. This equation
is known to describe a correlated random walk \cite{Ferrando159}
and, usually, is obtained phenomenologically by introducing
special relaxation flux terms to the original diffusion equation.
In \cite{myPRE} it was obtained rigorously by the Markovian
approximation \bea\label{Dcoh1}\nonumber &&\tilde
{D}_{coh}^{m}(z)=\frac{a^2}{4}\left(\frac{2 t_{inter}}{\hbar}
\right)^2\frac{ 1}{z+\tilde{\gamma}_{total}(z=0)}\\ &&\approx
\frac{a^2}{4}\left(\frac{2 t_{inter}}{\hbar}\right)^2\frac{
1}{z+\frac{1}{4}\Omega^2\tilde{\gamma}_c(z=0)}\eea for the
coherent part of the generalized diffusion coefficient and a
zero-width approximation for the Gaussian functions
(\ref{PhicApprox})-(\ref{PhipApprox}) (which define the incoherent
term) as a consequence of different time scales for both
mechanisms of dissipation.

Another remarkable feature of the short-time dynamics of the
adsorbate, described by Eq.~(\ref{difEqn}) along with
approximation (\ref{Dcoh1}) for the coherent contribution to the
generalized diffusion coefficient, follows from the expression for
a mean square displacement of the adparticle \bea\label{msd1}
&&\frac{\langle\Delta r(t)^2\rangle}{a^2}\!=\!\frac{2}{i\pi
a^2}\lim_{\epsilon\to
+0}\!\!\!\int\limits_{\epsilon-i\infty}^{\epsilon+i\infty}\!\frac{\tilde
D_{coh}(z)+\tilde D_{in}(z)}{z^2}\exp(z t) dz
\\\nonumber &&=\tilde{\gamma}_{inter}(0)t+\frac{8 t^2_{inter}\hbar^{-2}}{\tilde{\gamma}_{total}(0)
^2}\left[\exp(-\tilde{\gamma}_{total}(0)
t)-1+\tilde{\gamma}_{total}(0) t\right]. \eea While in the
hydrodynamic limit $t\to\infty$ Eq.~(\ref{msd1}) reproduces the
Einstein's law for the mean square displacement
\bea\label{msd-long} &&\frac{\langle\Delta
r(t)^2\rangle}{a^2}\!=\left(\tilde{\gamma}_{inter}(0)+\frac{8
t^2_{inter}\hbar^{-2}}{\tilde{\gamma}_{total}(0)} \right)t, \eea
in the short-time limit $t\ll 1/\tilde{\gamma}_{total}(0)$ the
second term of (\ref{msd1}) converts to $4 t^2_{inter}\hbar^{-2}
t^2$. The ballistic term, being obtained from the diffusion
equation for $f_{s,s}(t)$, is quite uncommon because in the
general case it appears only if one uses a Fokker-Planck equation
for the distribution function, depending on both a coordinate and
a velocity of the particle \cite{PhysA195,SurScience311}.

At the end of this Section we would like to point your attention
to the following matter. We call time-dependent diffusion
coefficients $D(t)$ the generalized diffusion ones, even though
this denotation is usually \cite{GCM4,Kubo} attributed to the
Laplace-transforms of $D(t)$. It should be stressed that the time
dependence of the kinetic kernels is much more informative than
frequency one: the generalized diffusion coefficient $D(t)$ is
directly related to the velocity autocorrelation function
$C_{s,s}^{JJ}(t)$, determined on the adsorption site $s$. The
investigation of its temporal behavior can help to visualize
the adparticle motion both at short and long times, and it is a
subject of Section VI. As for the next Section, we are going to
study the temperature dependence of the diffusion coefficients,
applying the Markovian approximation to $\tilde D_{coh}(z)$ and
$\tilde D_{in}(z)$.

\setcounter{equation}{0}
\section{Temperature behavior of the diffusion coefficients}

It is known that the experimentally measured values of the diffusion
coefficients $D_{exp}$ are usually associated with zeroth moments
of the generalized diffusion coefficients $D_0=\int_0^{\infty}D(t)
dt$, which is nothing but the Markovian approximation $D_{0}=\tilde
D(z=0)$ for their Laplace transforms. In fact, in experimental
conditions one deals with evaluation of the mean square
displacement  of the adparticle at times much larger than
$1/\tilde{\gamma}_{total}(0)$. Thus, measuring $\langle\Delta
r(t)^2\rangle$ one has to be sure that the influence of the
transient states is excluded, and duration of the atom-tracking
procedure is large enough to fall into the hydrodynamic region
$t\to\infty$. Otherwise, the value $D_{exp}$ will differ from its
theoretical prediction \cite{Ferrando}. We consider
the transition regimes of the adparticle motion in the next
Section. In this Section we investigate the temperature behavior
of the experimentally observed diffusion coefficients in the
framework of two-level dissipative model.

First of all, let us determine the conditions of the validity of
Markovian approximation. It is believed \cite{Ferrando,myPRE} that
memory effects can be neglected if the time scales describing the
adsorbate motion and those of the lattice dynamics are well
separated, $\omega_{max}/\Omega\gg 1$. Stronger
substrate-adsorbate coupling favors the Markovian approximation,
while the weak-coupling limit usually requires consideration of the
memory effects at the initial stage of the adparticle motion.

If the Markovian approximation is valid, the diffusion coefficient
is determined by the Einstein's law (\ref{msd-long}) as
\be\label{Dmarkov} \tilde
D(0)=\frac{a^2}{4}\left(\tilde{\gamma}_{inter}(0)+\frac{8
t^2_{inter}\hbar^{-2}}{\tilde{\gamma}_{total}(0)}\right). \ee

Taking into account the expressions (\ref{Tint}) for the renormalized
tunnelling amplitude and Eqs.~(\ref{Gamx})-(\ref{GamLLPlus}),
(\ref{Gamtotal}) for the kinetic kernels, and noting their time
dependence (\ref{PhicApprox})-(\ref{PhipApprox1}), it is easy to
perform an integration over $\tau$ and to obtain the final result.

Thus, in a strong-coupling limit $G\ge 0.1$ the diffusion
coefficient is completely defined by the incoherent term:
\bea\label{Dstrong} D_{strong}= \tilde
D_{in}(0)=\frac{a^2}{\omega_{max}}\left(\frac{t_1}{2\hbar}
\right)^2\sqrt{\frac{2\pi}{\eta_p}}\frac{\exp(-\eta_p/8 k_B
T)}{\sqrt{k_B T}}, \eea which, in its turn, is being determined by
the end-preserving processes.

In the weak-coupling limit $G\le 0.01$, and at a reasonable
assumption $t_1\ll\hbar\Omega$, the diffusion coefficient
\bea\label{Dweak}
D_{weak}=\frac{a^2}{\omega_{max}}\left(\frac{t_1}{2}
\right)^2\left[(\gamma_p+\gamma_c)+\frac{8\exp(-\varphi_p(0))}{\Omega^2\gamma_c}
\right] \eea is determined by both incoherent (the first term of
Eq.~(\ref{Dweak})) and coherent (the second term of
Eq.~(\ref{Dweak})) contributions. The end-changing $\gamma_c$ and
the end-preserv\-ing $\gamma_p$ rates can be presented in the
following form: \bea\label{gamC}
\gamma_c=\sqrt{\frac{\pi}{\eta_c|\ln\omega_0|}}\frac{\exp(-\eta_c|\ln\omega_0|/4
k_B T )}{\sqrt{k_B T}},\eea
\bea\label{gamP}&&\gamma_p=\omega_0^{2\eta_p k_B T-1}\left(
\frac{\omega_0^{(A-2)\eta_p k_B T}}{1-A \eta_p k_B T}-1 \right).
 \eea

It has to be noted that the expressions for $\gamma_c$ in both
limits coincide with those of Ref.~\cite{JCP2}, while the
expression for $\gamma_p$ in the weak-coupling limit differs from
the result of cited paper, which was obtained as the multiphonon
expansion of the corresponding end-preserving rates. Keeping in mind the
power law behavior (\ref{PhipApprox1}) of the end-preserving
kernel at weak-coupling, one can justify that a multiphonon
expansion is not valid in this particular case.

Now we have all necessary conditions to evaluate the temperature
behavior of the surface diffusion coefficients.
\begin{figure}[htb]
\centerline{\includegraphics[height=0.25\textheight]{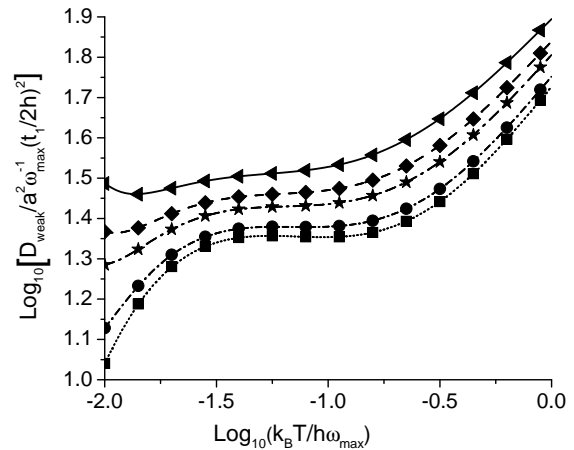}}
\vspace{0.1cm} \caption{Temperature dependence of the diffusion
coefficient in a weak-coupling limit. Model parameters:
$G=10^{-3}$, $\omega_0/\omega_{max}=5\times 10^{-4}$,
$t_1/\hbar\omega_{max}=10^{-5}$. Lines with triangles, diamonds,
stars, circles and squares denote, correspondingly, the values of
 $\Omega/\omega_{max}$=0.2, 0.25, 0.3,
0.5 and 1.} \label{Tslow}
\end{figure}
In Fig.~2 we present the log-log plot of the diffusion
coefficients in the weak-coupling limit as functions of temperature.
The sum of $D_{coh}$ and $D_{in}$ gives rise to quite a complex
temperature behavior of the overall diffusion coefficient, but for
all values of $\Omega$ shown, $D_{weak}$ is a relatively insensitive
function of the temperature.

The arguments about validity of the Markovian approximation at
values of $\Omega$, presented in Fig.~2, are not contradicting with
the condition $\omega_{max}/\Omega\gg 1$, as long as we are not
interested in the adsorbate motion at short time scales: according
to Eq.~(\ref{Dcoh1}), the zeroth moment of $D_{coh}(t)$ can be
calculated at any value of $\Omega$. We will see in the next
Section that it is not true if one investigates the intermediate
regimes $t\sim 1/\tilde{\gamma}_{total}(0)$, and it is necessary
to consider a diffusion equation which is nol-local in time.

In Fig.~3 we show the plot of the full diffusion coefficient at
stronger coupling to the lattice but still in the weak-coupling
regime.
\begin{figure}[htb]
\centerline{\includegraphics[height=0.25\textheight]{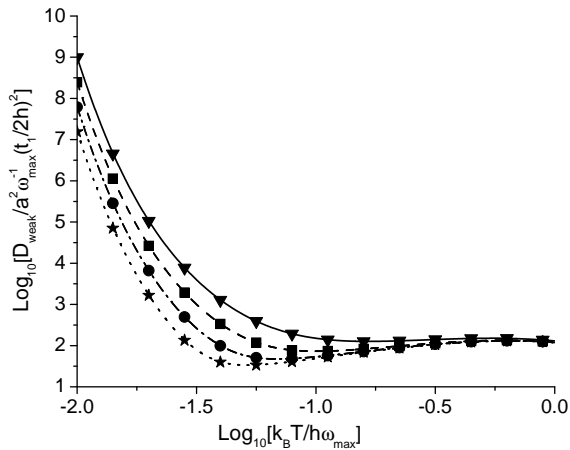}}
\vspace{0.6cm} \caption{Temperature dependence of the diffusion
coefficient in a weak-coupling limit. Model parameters:
$G=10^{-2}$, $\omega_0/\omega_{max}=5\times 10^{-4}$,
$t_1/\hbar\omega_{max}=10^{-5}$. Lines with triangles, squares,
circles and  stars denote, correspondingly, the values of
 $\Omega/\omega_{max}$=0.25, 0.5, 1,
and 2.} \label{Tstrong}
\end{figure}

\noindent The temperature dependence is quite different from that
shown in Fig.~2: at low temperatures the diffusion is dominated by
the coherent contribution, which is a strongly decreasing function
of $T$. According to (\ref{Dweak})-(\ref{gamC}), the coherent
contribution to the diffusion coefficient behaves as $\exp
(E_c/k_B T)$ at low temperatures and as $\sqrt{k_B T}$ at higher
temperatures. At high temperatures the main contribution comes
from the incoherent term, and $D_{weak}$ is a slowly increasing
function of $T$.

Let us briefly discuss the temperature behavior of $D_{weak}$ at low temperatures, shown in Fig.~3.
A qualitatively similar increase of the diffusion rate of H on Cu(001) below 20 K was observed experimentally in Ref.~\cite{single}.
However, in our case this behavior is just a result of the used two-level dissipative model, whereas in Ref.~\cite{single} it was attributed to the change of nonadiabatic response of the thermally excited electron-hole pairs to the diffusing particle.
When plotting Figs.~2 and 3 we just emphasize
that a crossover from one kind of temperature dependence
of the diffusion coefficient to another takes place even in a
relatively simple model. In the next Section we will show that
this crossover coincides perfectly (regarding to the coupling
constant $G$) with the change of the character of adparticle
dynamics at short times, when the memory effects have to be taken
into account.

\section{Transition regimes of the generalized diffusion coefficients}
\setcounter{equation}{0}

We have already mentioned that for visualization of the processes
of adparticle motion at short and intermediate times it is much
more convenient to perform an inverse Laplace transform of the
generalized diffusion coefficients according to
\bea\label{inverseL}\nonumber && D_{coh}(t)=\!\mbox{Re}\!\left[
\frac{(a t_{inter})^2}{2\pi i\hbar^2}\lim_{\epsilon\to
0}\!\!\int\limits_{\epsilon-i\infty}^{\epsilon+i\infty}\!\!
dz\exp(z
t)\frac{1}{z+\tilde{\gamma}_{total}(z)}\right] \\
&&=\left(\frac{a t_{inter}}{\hbar}\right)^2\mbox{Re}\left[
\sum_{i=1}^{\infty}\exp(-z_i
t)\frac{1}{1+\tilde{\gamma}'_{total}(z_i)}\right].\eea
The summation
in (\ref{inverseL}) in accordance with the residue theorem runs
over all poles $z_i$ of the integrand, which obey the condition
$\mbox{Re} [z_i]\le 0$. The expression (\ref{inverseL}) resembles
the results of the generalized collective modes theory
\cite{GCM1,GCM4}, postulating  the additive contribution of each
collective excitation to a particular time correlation function.
In our case, the summation is extended to the infinite number of
poles, and a main contribution comes from the terms with maximal
values of $\mbox{Re}[z_i]$ and weight factors
$[1+\tilde{\gamma}'_{total}(z_i)]^{-1}$. It has also to be
mentioned that the expression for $D_{coh}(t)$ can be even more
complicated if one deals with poles of the order $n>1$.

The expression for $D_{in}(t)$ follows immediately from the
definition (\ref{GamInter}) of the kernel
$\tilde{\gamma}_{inter}(z)$ and can be written down via the
end-changing/end-preserving functions
(\ref{PhicApprox})-(\ref{PhipApprox1}) as \bea\label{Din-t}
D_{in}(t)=(a t_{1}/h)^2\mbox{Re}\left[ \gamma_c(t)+\gamma_p(t)
\right]. \eea
We evaluate the generalized diffusion coefficients
in the weak-coupling regime when the memory effects are important
at the initial stage of the adparticle motion.
\begin{figure}[htb]
\centerline{\includegraphics[height=0.27\textheight,angle=0]{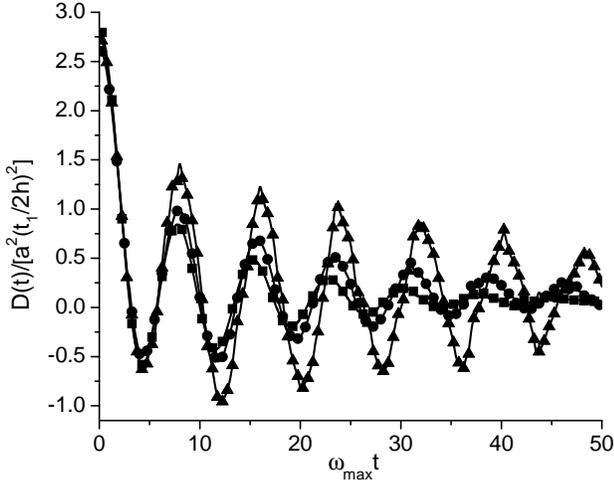}
}
 \vspace{1mm}
\caption{Generalized diffusion coefficients calculated at
$G=10^{-3}$, $t_1/\hbar\omega_{max}=10^{-5}$,
$\Omega/\omega_{max}=1$, and different temperatures: $k_B
T/\hbar\omega_{max}$=0.5 (triangles), 0.75 (circles) and 1
(squares).} \label{Dcoef}
\end{figure}
\noindent In Fig.~4 we present the time dependence of
$D(t)=D_{in}(t)+D_{coh}(t)$ at different temperatures. One can
draw some conclusions after observation of the plots.

First of all, we see a non-monotonic behavior of $D(t)$, and these
oscillations become more pronounced when the system temperature
decreases. At low temperatures thermal fluctuations of the lattice
are very small. Therefore, the lattice distortion caused by
interaction of the adsorbate with the phonon subsystem has no
time to relax, and the initial profile of the lattice potential
has no time to be restored after the particle passage. As a
result, the effective barrier is higher than its adiabatic value,
and the particle starts to oscillate being caged in the deformed
potential well. Such a behavior of $D(t)$ is observed even at the
temperatures comparable with $\hbar\omega_{max}/k_B$.

Secondly, at temperatures $k_B T/\hbar\omega_{max}<0.5$ (which are
not presented in Fig.~4) these oscillations persist on the time
scales, which are by two orders of magnitude higher than the
inverse Debye frequency. But even at higher temperatures there are
evident memory effects. In Ref.~\cite{myPRE} we calculated the
generalized diffusion coefficients at different values of $T$ and
$\Omega$. The general tendencies observed in \cite{myPRE} can be
formulated as follows: a low temperature $T$ and a high
vibrational frequency $\Omega$ favor the oscillation dynamics of
the particle, and so does a weak ``substrate--adsorbate''
interaction.

It would be interesting to relate this non-monotonic behavior of
$D(t)$ to a possible recrossing phenomenon \cite{JJ-TCF,TSWP,Japan}.
One can attribute the negative branches of $D(t)$ to the backward
motion of the adsorbate: the particle may cross the dividing
surface, located at the adsorption site $s$, due to the lattice
distortion that ``pushes'' the particle in the opposite direction
(with respect to that of the initial instant of motion). The
nature of oscillations of $D(t)$ is different from that of
$f_{s,s}(t)$, presented in Fig.~1. While in the coherent regime
the only reason of the non-monotonic behavior is an interplay
between the processes of transition and reflection of the wave
packet, associated with the adparticle, the non-monotonic
incoherent motion is determined by the adsorbate scattering on the
substrate atoms (one can verify that only $D_{coh}(t)$
contributes to the oscillatory adpartacle dynamics).

Now let us ask the question: what happens when one increases the
value of coupling constant $G$? An intuitive answer would state
that oscillations of $D(t)$ disappear at a moderate-to-strong
coupling. Indeed, at strong coupling, when the energy exchange
between the particle and the substrate atoms is faster, one can
apply the Markovian approximation (\ref{Dcoh1}) for the coherent
contribution to the generalized diffusion coefficient, which leads
to the exponential relaxation of $D_{coh}(t)$. However, fine
features of such transition regimes, when the character of the
adparticle motion changes from oscillatory to monotonic, need a
thorough analysis in the framework of the non-Markovian approach.

Thus, if one increases the coupling constant $G$ until the
oscillations of $D(t)$ disappear, one can obtain a critical value
$G_{cr}(T,\Omega)$ as a function of the temperature and
vibrational frequency, which separates two regimes of time
evolution: there is a plain relaxation of $D(t)$ at
$G>G_{cr}(T,\Omega)$, and a non-monotonic behavior at
$G<G_{cr}(T,\Omega)$. A detailed analysis shows that, at first, the
negative branch of $D(t)$ rises over the time axis, at that
oscillations of the generalized diffusion coefficients still
remain. So recrossing is vanishing, but the particle moves toward
the nearest adsorption site at one moment faster, at another
slower as if meeting obstacles.

At higher values of $G$ the oscillations completely disappear, and
the adsorbate motion is governed mainly by the incoherent term
$D_{in}(t)$. At $G\sim 0.1$ no coherent contribution is
evident. Moreover, in the expression (\ref{Din-t}) for $D_{in}(t)$
one has to use the strong-coupling limit for the end-preserving
kernel (\ref{PhipApprox}) rather than the weak-coupling form
(\ref{PhipApprox1}). At such values of $G$ the end-preserving rate
$\gamma_p(t)$ solely defines the adparticle dynamics, and the
temperature dependence of the diffusion coefficient is given by
Eq.~(\ref{Dstrong}).

\begin{figure}[htb]
\centerline{\includegraphics[height=0.3\textheight]{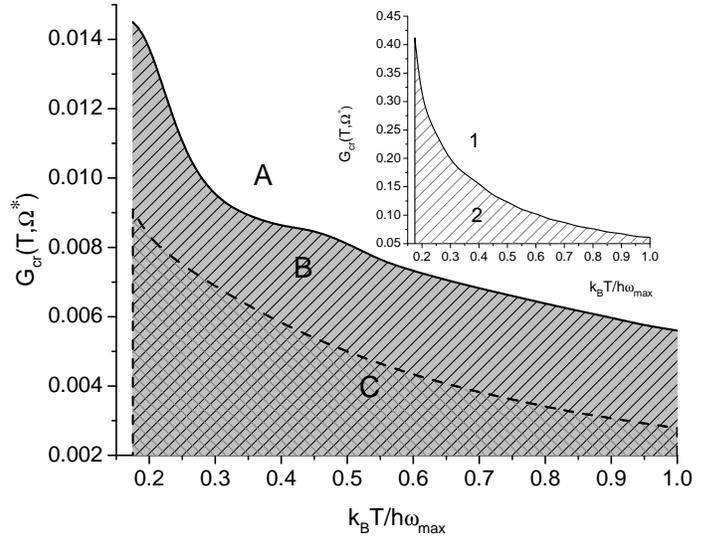}}
\vspace{0.1cm} \caption{The critical values of the coupling
constants $G_{cr}(T,\Omega^*)$ at $\Omega^*/\omega_{max}=1$, as
functions of temperature $T$, that separate domains of the
monotonic adparticle dynamics (A), oscillations without recrossing
(B), and an eventual recrossing (C). The inset shows the
temperature dependence of $G_{cr}(T,\Omega^*)$ that separates the
strong-coupling (1) and weak-coupling (2) domains.} \label{Gcr-T}
\end{figure}

\begin{figure}[htb]
\centerline{\includegraphics[height=0.3\textheight]{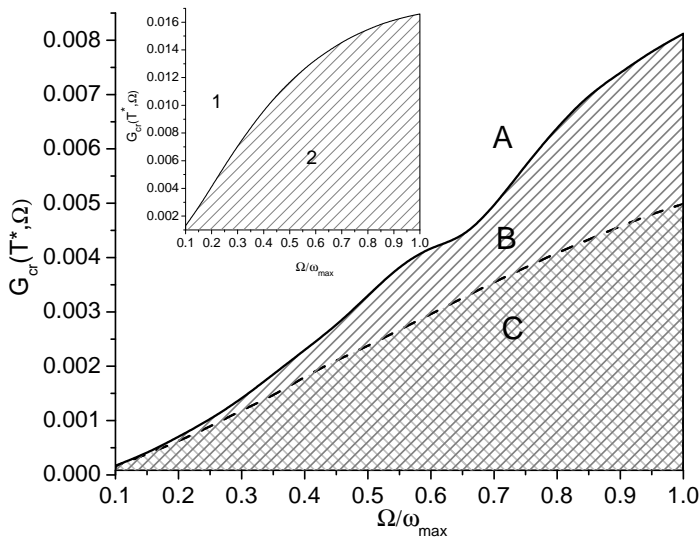}}
\vspace{0.1cm} \caption{The critical values of the coupling
constants $G_{cr}(T^*,\Omega)$ at $k_B T^*/\hbar\omega_{max}=0.5$,
as functions of the vibrational frequency $\Omega$, that separate
domains of the monotonic adparticle dynamics (A), oscillations
without recrossing (B), and an eventual recrossing (C). The inset
shows the frequency dependence of $G_{cr}(T^*,\Omega)$ that
separates domain (1), where $D_{coh}(t)$ decays as an exponential
function (see Eq.~(\ref{Dcoh1})), from that with essentially
non-Markovian behavior of $D_{coh}(t)$.} \label{Gcr-Omega}
\end{figure}

We present the above mentioned transition regimes of the
adparticle motion in Fig.~5 and 6 as plots of $G_{cr}$ vs.
temperature (at the fixed value of vibrational energy
$\Omega^*/\omega_{max}=1$, Fig.~5), and vs. vibrational energy (at
the fixed value of temperature $k_B T/\hbar\omega_{max}=0.5$,
Fig.~6). When inspecting these curves, one can observe a
remarkable feature: the transition domains of different dynamic
regimes of $D(t)$ with respect to the coupling constant $G$
coincide with the regions, where the temperature behavior of the
diffusion coefficients changes from a weakly dependent function of
$T$ to quite a sensitive function of temperature. Namely, the
region with a recrossing (the C-domain in Fig.~5) maps to a weakly
dependent temperature regime (see Fig.~2), and the region with
monotonic motion of the adparticle (the A-domain in Fig.~5)
corresponds to a strongly dependent temperature regime (see
Fig.~3). The domain B in Fig.~5, evidently, corresponds to
transition of the temperature behavior of experimentally measured
diffusion coefficient from weakly to strongly dependent function
of $T$.

The same tendencies can be traced in Fig.~6, where the plot of
$C_{cr}$ vs. $\Omega$ is presented at the fixed temperature. High
values of the vibrational frequency extend the domains (B and C)
of the non-monotonic adparticle dynamics. The insert in Fig.~6
shows that a transition from essentially non-Markovian dynamics to
the case, when $D_{coh}(t)$ decays as an exponential function (see
Eq.~(\ref{Dcoh1})), occurs in a moderate-to-strong coupling
domain. Of course, the most informative would be a
three-dimensional plot $G_{cr}(T,\Omega)$, but even such sections
of the critical coupling constants at fixed temperature and
vibrational energy, as in the above presented figures, give much food
for thought.

One of the assumptions inferred can be formulated as follows. If
the tendencies, presented in the ``critical diagrams'' in Figs.~5
and 6, remain in more sophisticated models describing quantum
surface diffusion, and a correspondence between the
``recrossing-monotonic motion'' transition, and the change in the
character of a $T$-dependence of the experimentally measured
diffusion coefficients  $D_{exp}$ is valid in general, then we
could give a prognosis about the temperature behavior of
$D_{exp}$, ha\-ving only the information about the dynamics of the
adsorbate at the initial stage of its motion. Otherwise, one has
to measure the mean square displacement of the particle at times
large enough to be sure that all remnants of the transition
regimes are excluded. As it has been already mentioned, due to
comparatively slow processes of quantum diffusion at low
temperatures (where the aforesaid transition regimes are the most
pronounced and durable), it would be useful to reduce the time of
atom-tracking experiment, because the longer action of the
measuring instrument, the greater influence on the system occurs.

It should be also noted that investigation of the transition
regimes in the framework of two-level dissipative models is a
topical problem \cite{Leggett}, and much efforts are put into a
study of the system crossover from one kind of its dynamics to
another. In a recent paper \cite{kytaj} the authors obtained the
``coherence--incoherence'' transition diagrams for the dissipative
two-level system with a nonzero bias and a sub-Ohmic bath as
functions of the power index $n$ in a low-frequency asymptotics
$J(\omega)\sim\omega^n$ of the spectral weight functions. Though
their results were obtained in zero-temperature limit, it would
be interesting to generalize such a model for $T\ne 0$ case to
investigate whether there is any other reason (except the
$G\to 0$ limit) for the ``coherence--decoherence'' transition.

At the end of the present Section we would like to touch upon the
study of the recrossing phenomenon once more. Though the considered
two-level dissipative model is too simplified to describe real
``substrate--adsorbate'' dynamics, especially in a comparison with
the models traditionally used during a direct evaluation of the
``flux-flux'' time correlation functions \cite{TSWP,Japan}, it is
worthy to compare them. On the one hand, in our model the lower
temperature, the more pronounced oscillations of the generalized
diffusion coefficient observed. This is different from the
results of Ref.~\cite{Japan} where high $T$ favors the recrossing
phenomenon. On the other hand, if we are in the low-temperature
weak-coupling domain and increase the value of $G$, we will
observe that there is a gradual transition from the domain C with
a pronounced recrossing to the domain B (see Figs.~5 and 6) with
oscillations but without recrossing. On the contrary, in the
high-temperature region with $k_B T/\hbar\omega_{max}\sim 1$ a
gradual increase of the coupling constant will give us a sudden
transition of the coherent part $D_{coh}(t)$ of the generalized
diffusion coefficient from the domain C to the region A of the
monotonic adparticle dynamics. However, this feature (not
presented graphically because we investigate the crossover for the
overall generalized diffusion coefficient $D(t)$ rather than for
$D_{coh}(t)$) is smeared out by the contribution of the incoherent
term, which has a ``long tail'' behavior (\ref{PhipApprox1}). If
one could eliminate these long tails, introducing additional
channels of the adparticle scattering, one would obtain an
interesting result: though the recrossing is less pronounced at high
$T$, it persists in a wider domain of coupling constant (the
region B completely vanishes). Such a behavior would resemble the
results of Ref.~\cite{Japan}. It has to be noted that the presented
model in the general features corresponds to the model F of the
cited paper (no surface motion is possible, both thermal
fluctuations and lattice distortions are permitted).

\section{Time dependence of the transition probabilities $f_{s,s+n}(t)$}
\setcounter{equation}{0}

In this section we study the off-diagonal non-equilibrium
distribution functions $f_{s,s+n}(t)$. There are some reasons to
look closer at the short-time dynamics of the  above mentioned
transition probabilities. First of all, the rate of the decay of
$f_{s,s'}(t)$ defines the time scales at which the contribution of
$D_{coh}(t)$ vanishes (let us re\-mind that to derive the
expression for the coherent term of the generalized diffusion
coefficients we solved Eq.~(\ref{fss1}) with respect to
$f_{s,s'}(t)$, and inserted the obtained result in the upper
equation (\ref{fss})). On the other hand, the off-diagonal
non-equilibrium distribution functions $f_{s,s+n}(t)$ can be
related to the time dependent ``flux--flux'' cross-correlation
functions, determined at the adjacent sites $s$ and $s+n$. These
time correlation functions are known to describe a multiple
crossing or a multi-hopping regime \cite{JJ-TCF,Japan}. The
multi-hopping facilitates an increase of the transport
coefficients, in contrast to the recrossing, which reduces the
total rate of the adparticle escape and lowers the value of the
diffusion coefficient.

We evaluate the non-equilibrium transition probabilities
$f_{s,s+n}(t)$, solving Eq.~(\ref{fss}) with respect to the
diagonal distribution functions $f_{s,s}(t)$ and inserting the
obtained result in Eq.~(\ref{fss1}). The final expressions for the
real and imaginary parts of $f_{s,s+n}(t)$ are \bea\label{RE}
\mbox{Re}[\tilde
f_{s,s+n}(z)]\!=\!\frac{f_{s,s+n}(0)}{z+\mbox{Re}\left[
\tilde{\gamma}_{inter}(z)+\tilde{\gamma}_{intra}(z)-\tilde{\gamma}_{LL}^+(z)
\right]}\!+\!o(t_1^3), \eea
\bea\label{IM} \mbox{Im}[\tilde
f_{s,s+n}(z)]=-\frac{t_{inter}}{\hbar}\frac{\tilde
f_{s+n,s+n}(z)-\tilde f_{s,s}(z)}{z+\mbox{Re}\left[
\tilde{\gamma}_{total}(z)\right]}+o(t_1^3), \eea where only linear
terms in the tunnelling amplitude $t_1$ are retained. It has to be
stressed that in the single particle limit there is no coupling
between the real and imaginary parts, while in the general case of
nonlinear kinetic equations $\mbox{Re}[\tilde f_{s,s+n}(z)]$ and
$\mbox{Im}[\tilde f_{s,s+n}(z)]$ are coupled to each other. The
second conclusion, which is to the point, is the following: the
decay rate of $\mbox{Re}[\tilde f_{s,s+n}(z)]$ is almost the same
as that of the coherent part $D_{coh}(t)$ of the generalized
diffusion coefficient,which is defined by the kernel
$\tilde{\gamma}_{total}(z)=\tilde{\gamma}_{inter}(z)+\tilde{\gamma}_{intra}(z)+\tilde{\gamma}_{LL}^+(z))$.
Obviously, in zero-coupling limit Eq.~(\ref{RE}) reproduces the
value of $f_{s,s+n}(t)$ which does not depend on time in the
coherent regime of motion, and the inverse Laplace transformation
of (\ref{IM}) in the limit $G\to 0$ gives us the expression
(\ref{nondiagonal}) for $\mbox{Im}[f_{s,s+n}(t)]$.

To study the time dependence of the transition probabilities we
perform a certain simplification of Eq.~(\ref{IM}). Namely, we
pass from the difference form in the numerator of (\ref{IM}) to
the derivative with respect to the adatom coordinate $r$. In fact,
we perform the continuous media approximation, similarly like it
was done in Ref.~\cite{myPRE}, when we obtained a Telegrapher's
equation, describing the correlated random motion of the adparticle.
After such a transformation, we are not bounded to a special
geometry of the lattice any more and can rewrite Eq.~(\ref{IM})
in the wave-vector--frequency representation for the imaginary part
of the transition probability $\tilde P(k,z)$ as follows:
\bea\label{Pk} \mbox{Im}[\tilde P(k,z)]=\frac{\hbar k}{2
t_{inter}}\frac{\mbox{Re}\left[\tilde
D_{coh}(z)\right]}{z+k^2\mbox{Re}\left[ \tilde D_{coh}(z)+\tilde
D_{in}(z)\right]}. \eea
Now, performing inverse Fourier and Laplace transformations, we can evaluate the $(r,t)$-dependence of the
imaginary part of the transition probability.
%\vspace {3mm}
\begin{figure}[htb]
\centerline{\includegraphics[height=0.26\textheight]{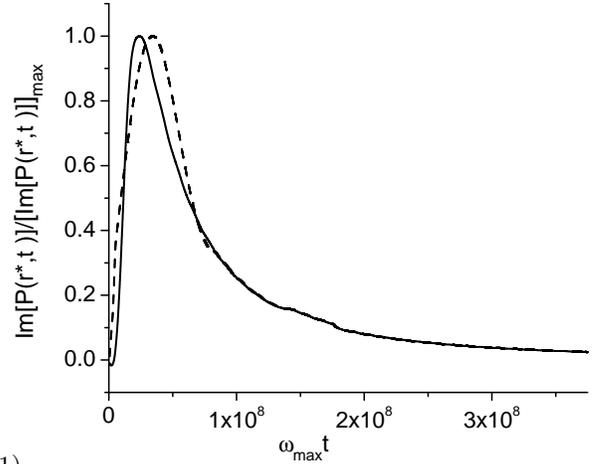}}
\vspace{0.2cm} \caption{Time dependence of the normalized function
$\mbox{Im}[P(r^*,t)]/[\mbox{Im}[P(r^*,t)]]_{max}$ at $r^*=r/a=1$,
$G=10^{-3}$, $\Omega/\omega_{max}=1$, $k_B T/\hbar\omega_{max}=1$,
and tunnelling amplitude $t_1/\hbar\omega_{max}=10^{-5}$. Solid
and dashed lines correspond to the Markovian and non-Markovian
approximations.}\label{P2markov}
\end{figure}
\noindent In Figs.~7 and 8 we present the time dependence of the
above mentioned functions (normalized at their maximum values) at
$r^*=r/a=1$. The evaluation is performed for two different values
of the tunnelling amplitude $t_1/\hbar\omega_{max}=10^{-5}$ and
$10^{-2}$. In the first case we observe that the adparticle
reaches the point $r^*$ at times of about 10$^7$ inverse Debye
frequency. The adparticle motion is very slow, the lattice has
enough time to relax at such huge times, and we do not observe any
noticeable difference between the Markovian approximation for
transition probability and the case, when the memory effects are
taken into account. \vspace{0.2cm}
\begin{figure}[htb]
\centerline{\includegraphics[height=0.29\textheight]{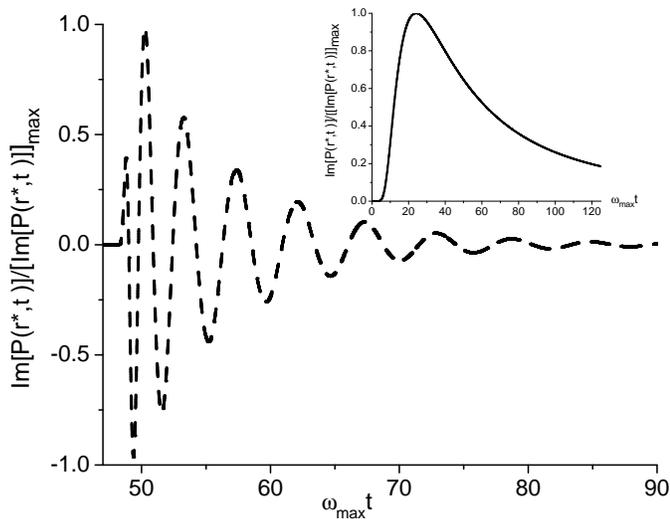}}
 \caption{Time dependence of the normalized function
$\mbox{Im}[P(r^*,t)]/[\mbox{Im}[P(r^*,t)]]_{max}$ at $r^*=r/a=1$,
$G=10^{-3}$, $\Omega/\omega_{max}=1$, $k_B T/\hbar\omega_{max}=1$,
and tunnelling amplitude $t_1/\hbar\omega_{max}=10^{-2}$. Solid
and dashed lines correspond to the Markovian and non-Markovian
approximations.}\label{P2nom-markov}
\end{figure}

A situation changes drastically (Fig.~8), when the value of
the tunnelling amplitude increases by three orders of magnitude. The
decay time of $P(r^*,t)$ is then comparable with that of the
generalized diffusion coefficient (see Fig.~4 for comparison). The
adparticle arrives at the point $r^*$ at times of about seven
Debye periods $2\pi/\omega_{max}$, when the lattice has not
relaxed completely, and the memory effects are still important.
Thus, a transition state that originally obstructs a multi-hop from
the site $s$ to the nearest neighboring one, is stabilized by the
lattice fluctuations: the height of the effective barrier at $s+n$
can be lower than its adiabatic value at some instant of time,
allowing the particle to perform a multiple crossing. We would
like to note the strongly aperiodic behavior of the transition
probability in contrast to the smooth relaxation in the Markovian
case. The negative branches of $P(r^*,t)$ mean that the ``transition
window'' for the multiple crossing is not always open, and at
other instants of time the effective barrier height can exceed its
static value, prohibiting multi-hops of the particle. On the other
hand, the width of oscillations growths in time while their
amplitude reduces. It means that multiple hops are less probable
when time increases, but the period favorable for them to proceed
becomes longer with increasing $t$.

We do not present plots for the real part of transition
probability, because its time behavior does not differ from that
of $D(t)$ (see Fig.~4). At the same time we have to emphasize that
all the curves in Figs.~7 and 8 correspond to the normalized
values of transition probabilities, while the non-normalized
values in the continuous media limit would be much smaller than
those of the distribution function $n(r^*,t)$ at the starting
point $r^*=0$ of the adparticle motion. In fact, $n(r^*,t)$ tends
to infinity at $r^*=0$, but even in the coherent limit and for a
discrete lattice model the curve for $\mbox{Im}[f_{0,1}(t)]$ lies
much lower (see Fig.~1) than that for $f_{0,0}(t)$. It is obvious
that the higher transition probability, the greater
contribution of the multi-hops to the overall diffusion
coefficient.

Thus, only rigorous evaluation of the site-dependent transition
probabilities $f_{s,s+n}(t)$ for a lattice with given geometry
would remove possible contradictions between the results, obtained
for the real discrete structure of the surface, and our model case
of a continuous media. Nevertheless, we believe that the basic
conclusions drawn in this Section are quite reliable in the
context of a qualitative analysis of the adparticle motion.

\section{Conclusions}

In this paper we made the systematic analysis of the dynamics of
the adparticle, which performs an underbarrier tunnelling from one
adsorption site to another as well as the vibrational transitions
between two levels inside a quantum well, when its motion changes
from coherent to incoherent due to interaction with acoustic
surface phonons. We investigated the short-time dynamics of the
particle in the case when its vibrational frequency $\Omega$ is
comparable with the Debye frequency $\omega_{max}$, and memory
effects have to be taken into account. Though the latter are
traditionally thought to be important only at transition regimes,
when the adparticle motion is neither ballistic nor purely
diffusive, and the experimentally measured diffusion coefficients
do not have to remember the details of intermediate period, we
showed that there is a close relation between the adparticle
dynamics at times $\tau\sim10\omega_{max}^{-1}$ and the
temperature dependence of diffusion coefficients. Namely, as the
coupling constant increases, the adparticle motion (initially
oscillatory) becomes more and more smooth indicating that the
temperature behavior of the diffusion coefficients $D(T)$ should
change from weakly dependent on $T$ to quite a sensitive function
of the temperature. Though we restricted ourselves by a
comparatively simple two-level dissipative model, we believe that
the above mentioned correspondence between the short-time dynamics
of the adsorbate and the temperature behavior of the diffusion
coefficients is valid for more sophisticated systems. If so, this
``$T$ vs. $t$'' correspondence could be helpful at experimental
evaluation of the diffusion coefficients because it would allow us
to give a prognosis about their temperature behavior, ha\-ving
only an information about the adatom dynamics at the initial stage
of its motion.

We showed that the coherent term of the generalized (time dependent)
diffusion coefficient, which is defined by the adparticle
scattering from the lattice, is responsible for the recrossing
phenomenon at weak-coupling, but at high-coupling regime its
contribution diminishes, and the particle motion is completely
determined by the incoherent term. We performed a quantitative
analysis of such transition regimes in terms of the critical
coupling constant $G_{cr}(T,\Omega)$ which depends on the
temperature and vibrational frequency. We compared the temperature
dependence of the recrossing with the results obtained by a direct
calculation of the ``velocity--velocity'' quantum time correlation
functions \cite{JJ-TCF,Japan}, and analyzed both common and
different features of the adparticle dynamics.

While the generalized diffusion coefficients are connected to the
``velocity--velocity'' autocorrelation functions, the transition
probabilities (non-diagonal distribution functions $f_{s,s'}(t)$)
can be related to the cross-correlation function
``velocity--velocity''. A study of the transition probabilities is
two-fold interesting: on the one hand, it shows how fast a loss of
the adparticle coherence occurs; on the other hand, it allows to
draw a conclusion about the phenomenon of multiple crossing. The
contribution of multiple jumps to the diffusion coefficient is
enhanced by the fact that in a double jump the random walker goes
a double distance. Since the diffusion coefficients is defined by
the squared jump length, the double jump can be important even if its
jump rate is comparatively smaller than that of a single jump.
%\vspace{-2mm}

We investigated the time dependence of the transition probabilities
in the model case of continuous media. It was shown that at very
small values $t_1/\hbar\omega_{max}=10^{-5}$ of the tunnelling
amplitude the real part of the transition probability (which defines
a multi-hop strength) decays at the same rate as the generalized
diffusion coefficients, while the particle approaches the nearest
adsorption site at much later times (which corresponds to the
maximum value of the imaginary part of $f_{s,s'}(t)$). The particle
moves very slow, the lattice has plenty time to relax, and the
Markovian approximation is quite applicable if one studies
non-equilibrium transition probabilities. Contrary, at
$t_1/\hbar\omega_{max}=10^{-2}$ the typical times of decay for
both real and imaginary parts of the transition probability are of
the same order, and remnants of the memory effects have an
influence on the multiple crossing, which is characterized by an
aperiodic oscillatory function. We also verified that the
zero-coupling limit reproduces the results for a coherent motion,
when the strength of transition $\mbox{Re}[f_{s,s+n}(t)]$ does not
change in time, and $\mbox{Im}[f_{s,s+n}(t)]$ is expressed
analytically via Bessel functions.

The memory effects could be said to preserve a particle coherence to a
certain extent: a portion of relaxation energy of the lattice is
delivered to the adparticles preventing them from thermalization
and maintaining the transition regimes from the coherent
(ballistic) motion to the incoherent (diffusive) one. This
reasoning would be even more realistic if one introduced the
additional channels of the adparticle scattering: an electronic
friction \cite{Kondo,Kagan}, a non-linear ``adsorbate--substrate''
interaction \cite{Zhu} or a direct ``adsorbate--adsorbate''
interaction \cite{Stasyuk}. First of all, this would solve the
``long-tails'' problem of the kinetic kernels \cite{MorozovBook}
and ensure their convergence without introduction of any
additional parameter like a particle life-time $\omega_0^{-1}$ in
a quantum well. On the other hand, additional interactions
introduce new typical time scales, which could be well separated
(the Markovian picture is then valid) or close to each other (then the
non-Markovian approach is necessary). Besides, by taking an
``adsorbate--adsorbate'' interaction into account one can go
beyond the limits of small coverage and study a concentration
dependence of the diffusion coefficients in addition to their
temperature behavior \cite{Cdepend}. We believe that all these
directions are quite interesting from a viewpoint of the study of
transition regimes of the adsorbate and could be the subject of
future investigations.

\section*{Acknowledgement}
This work was partially supported by the Project ``Models of the
quantum-statistical description of catalytic processes at the
metallic surfaces''(Lviv Polytechnic National University), No.
0110U001091.

\section*{Appendix}
\renewcommand{\theequation}{A.\arabic{equation}}
\setcounter{equation}{0}

To solve Eqs.~(\ref{fssCoh})-(\ref{fssCoh1}) let us perform at
first a Fourier transformation for the creation/annihilation
operators \be\label{Fourier} a^{\dagger}_s=\frac{1}{\sqrt
N}\sum\limits_k\exp(i k s) a^{\dagger}_k,\quad a_s=\frac{1}{\sqrt
N}\sum\limits_k\exp(-i k s) a_k, \ee passing from the site
representation $a^{\dagger}_s$, $a_s$ to wave-vector
representation $a^{\dagger}_k$, $a_k$ with $k=(2\pi/N) m$
[$k=(2\pi/N) (m+1/2)$] for the lattice with the odd [even] numbers
of adsorption sites and $m=-N/2,-N/2+1,\ldots,N/2-1$ [if $N$ is
even] or $m=-(N-1)/2, -(N-1)/2+1,\ldots,(N-1)/2$ [if $N$ is odd].

Inserting (\ref{Fourier}) in Eqs.~(\ref{fssCoh})-(\ref{fssCoh1}) for the coherent motion, we
obtain the following equation for the intermediate
distribution function $F_{k,k'}(t)\equiv\langle a^{\dagger}_{k'}
a_k\rangle^t$: \be\label{DFkk1} \frac{\partial
F_{k,k'}(t)}{\partial t}=-\frac{2 i t_0}{\hbar}(\cos k-\cos
k')F_{k,k'}(t),\ee which is easily solved, giving
\be\label{Fkk1} F_{k,k'}(t)=\exp\left[\frac{-2 i t_0}{\hbar}(\cos k-\cos
k')t\right].\ee
In the infinite lattice limit $N\to\infty$, we can pass from summation over $k$ to
integration over continuous wave-vector according to
$\frac{1}{\sqrt N}\sum_k\cdots \longrightarrow
\frac{1}{2\pi}\int^{\pi}_{-\pi}\cdots dk$. After that we obtain an
integral representation for the diagonal one-particle
non-equilibrium distribution function as follows:
\bea\label{J0J0}&&\nonumber
f_{s,s}(\tau)=\frac{1}{4\pi^2}\left|\int\limits^{\pi}_{-\pi}d
k[\cos(k s)+i\sin(k s)]\right.\\
&&\left.\times\exp\left[\frac{-2 i t_0}{\hbar} \tau\cos
k\right] \right|^2=J_s^2(2 t_0\hbar^{-1} \tau), \eea which is nothing but the
squared s-th order Bessel function. Similarly, we can obtain the
expression (\ref{nondiagonal}) for the imaginary part of the
transition probability $f_{s,s+1}(t)$.

To show that the last term in Eq.~(\ref{difEqn}) does not
contribute to the mean square displacement $\langle\Delta
r(t)^2\rangle=a^2\sum\limits_{s=1}^N s^2 f_{s,s}(t)$ of the
particle let us perform for simplicity the Markovian approximation
for the generalized diffusion equation (\ref{difEqn}). Using
Fourier transformation (\ref{Fourier}) we obtain the evolution
equation for the intermediate distribution function in the
following form: \bea\label{Fkk-msd}\nonumber &&\frac{\partial
F_{kk'}(t)}{\partial t}=-\frac{F_{kk'}(t)}{a^2}\left\{2[\tilde
D_{coh}(0)+\tilde D_{in}(0)][1-\cos(k-k')]\right.
\\
&&+\left.\tilde D_{coh}(0)[\cos 2k+\cos
2k'-2\cos(k+k')]\right\},\eea where the second term in the braces
is related to the last term in the r.h.s of Eq.~(\ref{difEqn}),
which involves the ``long distance'' transition probabilities
$f_{s\pm n,s\mp n}(t)$, $f_{s,s\mp 2n}(t)$.

The evolution equation for the mean square displacement can be
written down as follows: \bea\label{t-msd}\nonumber &&
\frac{d\langle\Delta r(t)^2\rangle}{d
t}=a^2\sum\limits_{s=1}^N\sum\limits_{k,q}s^2\exp(i q s)\dot
F_{k,k-q}(t)\\
&&=-a^2\sum\limits_{k,q}\dot
F_{k,k-q}(t)\frac{d^2}{dq^2}\left(\sum\limits_{s=1}^N\exp(i q
s)\right).\eea
Noting that the sum in the brackets yields $N$
times Kronecker delta-symbol, which in the infinite lattice limit
converts to Dirac delta-function $\delta(q)$, and integrating
(\ref{t-msd}) by parts with taking into account (\ref{Fkk-msd}),
one can verify that only the first term in braces of
Eq.~(\ref{Fkk-msd}) contributes to the evolution equation for
$\langle\Delta r(t)^2\rangle$: \bea\label{1st}\nonumber
&&\frac{d\langle\Delta r(t)^2\rangle}{d t}=[\tilde
D_{coh}(0)+\tilde
D_{in}(0)]/\pi\int\limits_{-\pi}^{\pi}dk\\
&& \times\!\!\!\int\limits_{-\pi}^{\pi}\!dq[1-\cos
q]F_{k,k-q}(t)\frac{d^2}{dq^2}\delta(q)=2[\tilde D_{coh}(0)+\tilde
D_{in}(0)],\eea
while the second term vanishes at the integration
over $k$:
%\\[-14ex]
\bea\label{2nd}\nonumber &&
\int\limits_{-\pi}^{\pi}dk\int\limits_{-\pi}^{\pi}dq\left\{
\cos(2k)+\cos(k-q)-2\cos(2k-q) \right\}\\
&&\times F_{k,k-q}(t)
\frac{d^2}{dq^2}\delta(q)=\!\!\int\limits_{-\pi}^{\pi}\!\!dk
[2\cos(2k)-\cos k]F_{k,k}(t)=0,\eea because $F_{k,k}(t)\equiv 1$,
and all derivatives of $F_{k,k}(t)$ with respect to wave-vector
vanish.

The presented above calculation can be generalized to the 2$D$
case, or to the case when the memory effects are taken into
account. As for the 2$D$ coherent regime, the only modification of
Eqs.~(\ref{diagonal})-(\ref{nondiagonal}) consists in doubling of
the power indexes at Bessel functions.


\begin{thebibliography}{99}
%1
\bibitem{catalysis} {\it Hydrogen Effects in Catalysis}, edit. by Z. Paal and P.~G.~Menon ~Dekker,
(New York, 1988).
%2
\bibitem{q-jumps}
P.~Costamagna, S.~Srinivasan, Journ. Power Sourc. {\bf 102}, 242
(2001).
%3
\bibitem{Htransfer} L.J.~Lauhon and W.~Ho, J. Phys. Chem. {\bf 105}, 3987 (2000).
%4
\bibitem{PRB-41-1990}
I.C.~da Cunha Lima, A.~Troper, and S.C.~Ying, Phys.~Rev.~B {\bf
41}, 11798 (1990).
%5
\bibitem{qkinet}
V.~Pouthier, J.~C.~Light, Journ. Chem. Phys. {\bf 133},  1204
(2000).
%6
\bibitem{Ferrando}
T.~Ala-Nissilayz, R.~Ferrando, and S.C.~Ying, Advances in Physics
{\bf 51}, No.~3, 949 (2002).
%7
\bibitem{JJ-TCF}
G.~Wahnstr\"om, K.~Haug, and H.~Metiu, Chem. Phys. Lett. {\bf
148}, 158 (1988).
%8
\bibitem{Japan}
T.~Taniike, and K.~Yamashita, Chem. Phys. {\bf 304}, 159 (2004).
%9
\bibitem{single}
W.~Ho, Journ. Chem. Phys. {\bf 117}, 11033 (2002).
%10
\bibitem{qmd}R.~Baer, Y.~Zeiri and R.~Kosloff, Surf. Sci. {\bf
411},  L783 (1998)
%11
\bibitem{36Ferrando}
 S.C.~Badescu, S.C.~Ying, and T.~Ala-Nissila,  Phys. Rev. Lett. {\bf 86}, 5092 (2001).
%12
\bibitem{Ohresser}
P.~Ohresser, H.~Bulou, S.S.~Dhesi, C.~Boeglin, B.~Lazarovits,
E.~Gaudry, I.~Chado, J.~Faerber, and F.~Scheurer, Phys. Rev. Lett.
{\bf 95}, 195901 (2005).
%13
\bibitem{Casado}
R. Mart\'inez-Casado, A. Sanz, and S. Miret-Art\'es, J. Chem.
Phys. {\bf 129}, 184704 (2008).
%14
\bibitem{Kondo}
J.~Kondo, Physica {\bf 125B}, 279 (1984).
%15
\bibitem{PRL-62-1989}
R.F.~Kiefl, R.~Kadono, J.H.~Brewer, G.M.~Luke, H.K.~Yen, M.~Celio,
and E.J.~Ansaldo, Phys. Rev. Lett. {\bf 62}, 792 (1989).
%16
%\bibitem{PRB-39-1989}
%R.~Kadono, J.~Imazato, T.~Matsuzaki, K.~Nishiyama, K. Nagamine, T.
%Yamazaki, D.~Richter, and J.-M.~Welter, Phys. Rev. B {\bf 39}, 23
%(1989).
%17
\bibitem{18PreprintMorozov}
Yu.~Kagan, N.V.~Prokofiev, Zh. Eksp. Teor. Fiz. {\bf 96}, 2209
(1989) [Sov. Phys. JETP. {\bf 69}, 1250 (1989)].
%18
\bibitem{JCP1} P.D.~Reilly, R.A.~Harris, and K.B.~Whaley, Journ. Chem. Phys. {\bf 95},  8599
(1991).
%19
\bibitem{JCP2} P.D.~Reilly, R.A.~Harris, and K.B.~Whaley, Journ. Chem. Phys. {\bf 97},  6975
(1992).
%20
\bibitem{Zhu} X.~D.~Zhu and L.~Deng, Phys. Rev. B {\bf 48}, 17527
(1993).
%21
\bibitem{Kagan}
Yu.~Kagan, N.~V.Prokofiev, Zh. Eksp. Teor. Fiz. {\bf 90}, 2176
(1986) [Sov. Phys. JETP. {\bf 63}, 1276 (1986)].
%22
%\nopagebreak[3]
\bibitem{Stasyuk}
W.~Brenig, Surf. Sci. {\bf 291}, 207 (1993).
%23
\bibitem{MorozovBook}
 D.N.~Zubarev, V.G.~Morozov, G.~R\"opke, \it{Statistical Mechanics of
Nonequilibrium Processes
%Vol.2, Relaxation and Hydrodynamic
%Processes
}\rm, (Fizmatlit, Moscow, 2002, in Russian).
%24
%\bibitem{MorozovPreprint}
%V.G.~Morozov, Unpublished.
%25
\bibitem{PhysA195}
R.~Ferrando, R.~Spadacini, G.E.~Tommei, and G. Caratti, Physica A
{\bf 195}, 506 (1993).
%26
\bibitem{SurScience311}
R.~Ferrando, R.~Spadacini, G.E.~Tommei, and G. Caratti, Surface
Science, {\bf 311}, 411 (1994).
%27
%\bibitem{PRE-51-1995}
%E.~Pollak, and P.~Talkner, Phys. Rev. E {\bf 51}, 1868 (1995).
%
\bibitem{TSWP}
D.H.~Zhang, J.C.~Light, and Soo-Y. Lee, Journ. Chem. Phys. {\bf
111}, 5741 (1999).
%28
%\bibitem{Kramers}
%P.~H\"anggi, P.~Talkner, M.~Borkovec, Rev. Mod. Phys. {\bf 62},
%251 (1990).
%29
%
%30
\bibitem{myPRE}
V.~V.~Ignatyuk, Phys. Rev. E {\bf 80}, 041133 (2009).
%31
\bibitem{cmp2004} V~.V~.Ignatyuk, V.~G~.Morozov,
Condens. Matter Phys. {\bf 7}, No 3(39), 579 (2004).
%32
\bibitem{inclusion}
P.D.~Reilly, R.A.~Harris, and K.B.~Whaley, Phys. Rev. B {\bf 47},
5721 (1993).
%37
\bibitem{Leggett} A.J.~Leggett, S.~Chakravarty, A.T.~Dorsey,
M.P.A.~Fisher, A.~Garg, and W.~Zwerger, Rev. Mod. Phys. {\bf 59},
1 (1987).
\bibitem{GCM1}
I.M.~Mryglod, I.P.~Omelyan, and M.V.~Tokarchuk, Mol. Phys. {\bf
84} 235, (1995).
%33
%\bibitem{GCM2}
%I.M.~Mryglod, I.P.~Omelyan, and R.~Folk, J. Phys.: Cond. Matt.
%{\bf 15}, S83 (2003).
%%34
%\bibitem{GCM3}
%S.~Cazzato, T.~Scopigno, T.~Bryk, I.~Mryglod, and G.~Ruocco, Phys.
%Rev. B. {\bf 77}, 094204: 1-6 (2008).
%%35
\bibitem{GCM4}
V.V.~Ignatyuk, I.M.~Mryglod, and M.V.~Tokarchuk, Low Temp. Phys.
{\bf 25}, 857 (1999).
%36
%38
\bibitem{Chvoj}
Z.~Chvoj, J. Phys.: Cond. Matt. {\bf 12}, 2135 (2000).
%39
\bibitem{Ferrando159}
D.~Dou, J.~Casas-Vazquez, and G.~Lebon, \it{Extended Irreversible
Thermodynamics}\rm\, (Springer, Berlin, 1998).
%
%40
\bibitem{Kubo}
R.~Kubo, M.~Toda and N.~Hashitsume, {\it Statistical Physics} II
(Springer, Berlin, 1991).
%41
\bibitem{kytaj}
Congjun Gan and Hang Zheng, Phys. Rev. E {\bf 80}, 041106 (2009).
%
\bibitem{Cdepend} A.~Wong, A.~Lee, and X.D.~Zhu, Phys. Rev. B {\bf 51}, 4418 (1995).
%
%\bibitem{pollak}
%S.~Miret-Art\'es, and E.~Pollak, J.Phys.: Condens. Matter {\bf
%15}, S4133 (2005).
%%2
%%3
%%4
%%\bibitem{prl94-086101}
%%Yu.~Jia, W.~Zhu, E.G.~Wang, Y.~Huo, and Z.~Zhang, Phys. Rev. Lett.
%%{\bf 94}, 086101 (2005).
%%5
%\bibitem{preprint}
%R.~Mart\'inez-Casado, A.S.~Sanz, and S.~Miret-Art\'es, arXiv:
%cond-mat/0803.0535v1 (2008).
%%6
%%7
%%8
%%9
%%8
%%\bibitem{MorozovBookR}
%% D.N.~Zubarev, V.G.~Morozov, G.~R\"opke, \it{Statistical Mechanics of
%%Nonequilibrium Processes. Vol.1}\rm, (Fizmatlit, Moscow, 2002) (in
%%Russian).
%%10
%%11
%%12
%%13
%
%%14
%
%%15
%
%%16
%%17
%%18
%\bibitem{CO}
%K.A.~Fichthorn, P.~G. Balan, Journ. Chem. Phys. {\bf 101}, 10028
%(1994).
%%19
%\bibitem{my2003}
%P.P.~Kostrobii, Yu.K.~Rudavskii, V.V.~Ignatyuk, and
%M.V.~Tokarchuk, Condens. Matter Phys. {\bf 6}, No.~3(35), 409
%(2003).
%%20
%%22
%\bibitem{16PM}
%Yu.~Kagan, M.I.~Klinger, Journ. Phys. C {\bf 7}, 2791 (1974).
%%23
%%24
%%
%%
%\bibitem{we}
%V.V.~Ignatyuk, M.V.~Tokarchuk, P.P.~Kostrobij, Condens. Matter
%Phys. {\bf 9}, No.~1(45), 55 (2006).
%%
%%
%%
%%28
%%the end of Introduction
%%30
%\bibitem{15JCP2}
%T.~Holstein, Ann. Phys. {\bf 8}, 325 (1959).
%%31
%\bibitem{16JCP2}
%C.P.~Flynn, and A.M.~Stoneham, Phys. Rev. B {\bf 1}, 3966 (1970).
%%32
%\bibitem{421Ferrando}
%T.R.~Mattson, and G.~Wahnstr\"om, Phys. Rev. B {\bf 56}, 14944
%(1997).
%%33
%\bibitem{428Ferrando}
%H.~Okuyama, T.~Ueda , T.~Aruga, and M.~Nishijima, Phys. Rev. B
%{\bf 63}, 233403 (2001).
%%34
%Lett. {\bf 86}, 5092 (2001).
%%35
%\bibitem{15PreprintMorozov}
%Yu.~Kagan, L.A.~Maksimov, Zh. Eksp. Teor. Fiz. {\bf 65}, 622
%(1973) [Sov. Phys. JETP {\bf 38}, 307 (1974)].
%36
%
%
%\bibitem{single164}
%A.P.~Graham, A.~Menzel, and J.P.~Toennies, Journ. Chem. Phys. {\bf
%111}, 1676 (1999).
%%15
%\bibitem{Ohresser}
%P.~Ohresser, H.~Bulou, S.S.~Dhesi, C.~Boeglin, B.~Lazarovits,
%E.~Gaudry, I.~Chado, J.~Faerber, and F.~Scheurer, Phys. Rev. Lett.
%{\bf 95}, 195901 (2005).
%



\end{thebibliography}
\end{document}